\documentclass[aps,prd,superscriptaddress,twocolumn,nofootinbib]{revtex4-1}
\usepackage[utf8]{inputenc}
\usepackage{dsfont}
\usepackage{amsfonts}
\usepackage{amsmath}
\allowdisplaybreaks[4]        
\usepackage{amssymb}
\usepackage{euscript}     
\usepackage{braket}
\usepackage{starfont}
\usepackage{color,soul}         
\usepackage{tensor}        
\usepackage{amsthm}
\usepackage{graphicx}
\usepackage{slashed}
\usepackage{leftidx}
\usepackage{subfigure}
\usepackage{bbm}
\definecolor{outerspace}{rgb}{0.25, 0.29, 0.3}
\definecolor{scarlet}{rgb}{1.0, 0.13, 0.0}
\usepackage[header,title,page,titletoc]{appendix}  
\definecolor{princetonorange}{rgb}{1.0, 0.56, 0.0}
\definecolor{WildStrawberry}{rgb}{1.0, 0.26, 0.64}
\definecolor{rossocorsa}{rgb}{0.83, 0.0, 0.0}
\definecolor{navyblue}{rgb}{0.0, 0.0, 0.5}
\usepackage{float}
\usepackage[paper=letterpaper,margin=1in]{geometry}
\usepackage{mathtools}

\newcommand{\req}[1]{(\ref{#1})} 

\newcommand{\bea}{\begin{eqnarray}}

\newcommand{\eea}{\end{eqnarray}}
\newcommand{\ba}{\begin{eqnarray}}
\newcommand{\ea}{\end{eqnarray}}

\newcommand{\be}{\begin{equation}}
\newcommand{\ee}{\end{equation} }
\newcommand{\beqa}{\begin{eqnarray}}
\newcommand{\eeqa}{\end{eqnarray}}
\newcommand{\beqar}{\begin{eqnarray*}}
\newcommand{\eeqar}{\end{eqnarray*}}

\renewcommand{\req}[1]{(\ref{#1})}

\makeatletter
\newcommand{\dal}{\mathop{\mathpalette\dal@\relax}}
\newcommand{\dal@}[2]{%
  \begingroup
  \sbox\z@{$\m@th#1\square$}%
  \dimen0=\fontdimen8
    \ifx#1\displaystyle\textfont\else
    \ifx#1\textstyle\textfont\else
    \ifx#1\scriptstyle\scriptfont\else
    \scriptscriptfont\fi\fi\fi3
  \makebox[\wd\z@]{%
    \hbox to \ht\z@{%
      \vrule width \dimen0
      \kern-\dimen0
      \vbox to \ht\z@{
        \hrule height \dimen0 width \ht\z@
        \vss
        \hrule height 2\dimen0
      }%
      \kern-2.5\dimen0
      \vrule width 2.5\dimen0
    }%
  }%
  \endgroup
}
\makeatother

\usepackage{hyperref}
\hypersetup{
    colorlinks,
    citecolor=blue,
    filecolor=navyblue,
    linkcolor=navyblue,
    urlcolor=navyblue
}

\begin{document}

\title{Boson stars with nonlinear sigma models}
\author{Pablo A. Cano}
\email{pablo.cano@icc.ub.edu}
\affiliation{Departament de F\'isica Qu\`antica i Astrof\'isica, Institut de Ci\`encies del Cosmos\\
 Universitat de Barcelona, Mart\'i i Franqu\`es 1, E-08028 Barcelona, Spain }
\affiliation{Instituut voor Theoretische Fysica, KU Leuven. Celestijnenlaan 200D, B-3001 Leuven, Belgium. }

\author{Ludovico Machet}
\email{ludovico.machet@kuleuven.be}
\affiliation{Instituut voor Theoretische Fysica, KU Leuven. Celestijnenlaan 200D, B-3001 Leuven, Belgium. }
\affiliation{Physique Théorique et Mathématique, ULB, Boulevard du Triomphe, B-1050 Bruxelles, Belgium}

\author{Charlotte Myin}
\email{charlotte.myin@ds.mpg.de}
\affiliation{Instituut voor Theoretische Fysica, KU Leuven. Celestijnenlaan 200D, B-3001 Leuven, Belgium. }


\begin{abstract}
Boson stars arise as solutions of a massive complex scalar field coupled to gravity. 
A variety of scalar potentials,  giving rise to different types of boson stars, have been studied in the literature.  Here we study instead the effect of promoting the kinetic term of the scalar field to a nonlinear sigma model --- an extension that is naturally motivated by UV completions of gravity like string theory. 
We consider the $\mathrm{O}(3)$ and $\mathrm{SL}(2,\mathbb{R})$ sigma models with minimally interacting potentials and obtain their boson star solutions. We study the maximum mass and compactness of the solutions as a function of the curvature of the sigma model and compare the results to the prototypical case of mini boson stars, which are recovered in the case of vanishing curvature. The effect of the curvature turns out to be dramatic. 
While $\mathrm{O}(3)$ stars are massive and can reach a size as compact as $R\sim 3.3 GM$,  $\mathrm{SL}(2,\mathbb{R})$ stars are much more diffuse and only astrophysically viable if the bosons are ultralight. 
These results show that the scalar kinetic term is at least as important as the potential in determining the properties of boson stars.

\end{abstract}
\maketitle

\section{Introduction}
The study of gravitating particle-like objects --- self-gravitating solutions without a horizon or a singularity --- forms an important chapter in the history of general relativity, dating back to Wheeler's geons \cite{PhysRev.97.511,RevModPhys.29.480}.  Today, at the dawn of the era of gravitational wave astronomy \cite{LIGOScientific:2016aoc}, the interest in this type of objects has spiked. Indeed, among the many possibilities these experiments offer us is the search for (horizonless) exotic compact objects, alternative to black holes, which would produce characteristic gravitational wave signatures \cite{Cardoso:2017cqb,Cardoso:2019rvt}.  Many candidates for exotic compact objects have been proposed in the literature, \textit{e.g.} \cite{Morris:1988tu,Mazur:2001fv,Mathur:2005zp,Damour:2007ap,Holdom:2016nek,Bueno:2017hyj,Mayerson:2023wck}, but among them, boson stars \cite{JETZER1992163, Liebling:2012fv, Visinelli:2021uve} stand out as the best motivated ones, as they have a robust dynamical foundation \cite{Lee:1988av,Gleiser:1988rq} and they could be part of the dark matter content of the universe \cite{JETZER1992163, Schunck:2003kk, Macedo:2013qea}.
 
Boson stars are self-gravitating condensates of massive bosons prevented from collapsing by the Heisenberg uncertainty principle. 
These objects arise as solutions of general relativity coupled to massive bosonic fields (either scalar or Proca fields) with a continuous symmetry.  The existence of this symmetry allows these fields to generate a stationary stress-energy tensor and a conserved charge by rotating in field space. 

The simplest theory leading to boson stars corresponds to a complex scalar field $\Phi$ invariant under phase shifts $\Phi\rightarrow e^{i \alpha} \Phi$, with a Lagrangian of the form 
\begin{equation}\label{eq:SM}
\mathcal{L}_{\Phi}=-\frac{1}{2}\partial_{\mu}\Phi\partial^{\mu}\bar{\Phi}-V\left(|\Phi|^2\right)\, .
\end{equation}
This kind of models, with a variety of potentials $V\left(|\Phi|^2\right)$, have been extensively studied in the literature. 
The properties of boson stars change widely depending on the potential, and some of the most studied models are mini boson stars \cite{PhysRev.172.1331,PhysRev.187.1767} --- with a non-interactive potential --- massive boson stars \cite{PhysRevLett.57.2485} --- including a quartic self-interaction \cite{Khlopov:1985jw}--- solitonic boson stars \cite{PhysRevD.35.3637, PhysRevD.35.3658, LEE1992251} --- which unlike the others, persist in flat spacetime --- or axion boson stars \cite{Guerra:2019srj, Delgado:2020udb} --- with a periodic potential inspired by QCD. Other related solutions are the recently proposed $\ell$-boson stars that arise in a system with $N$ complex Klein-Gordon fields \cite{Alcubierre:2018ahf,Alcubierre:2019qnh,Jaramillo:2020rsv,Alcubierre:2021psa}. It would be impossible to list here all the existing boson star models, so we refer to \cite{Liebling:2012fv, Mielke:2016war, Visinelli:2021uve} and the references therein for a comprehensive review. 

Despite this zoo of models,  one can make the case that there are interesting and natural extensions of this family of theories yet to be explored. In fact, instead of just changing the potential, one could more generally promote \req{eq:SM} to a nonlinear sigma model
\begin{equation}
\mathcal{L}_{\phi}=-\frac{1}{2}G_{AB}(\phi)\partial_{\mu}\phi^{A}\partial^{\mu}\phi^{B}-V\left(\phi^{A}\right)\, .
\end{equation}
Here, the scalars $\phi^{A}$ (with $A=1,\ldots, N$) can be regarded as coordinates in a target space: the scalar manifold. The kinetic matrix $G_{AB}(\phi)$ is the metric in this manifold, which in general can be curved. These models, too, can exhibit continuous symmetries required for the construction of boson stars. This happens whenever  the scalar manifold possesses an isometry. In fact, this kind of models are commonplace in supergravity and compactified string theory effective actions, where the scalar manifold is very often a symmetric space \cite{Ortin:2015hya}. Also in those theories, a potential $V(\phi^{A})$ generally arises when the symmetries are gauged.  Thus, this is a very natural extension of \req{eq:SM} motivated by high-energy physics. This is even more significative taking into account that the scalar field can typically reach Planckian values in a boson star \cite{Ho:1999hs}, and hence one would be probing the theory all the way up to Planck scale, where these nonlinearities become relevant.

The goal of this paper is to show the existence and study the properties of boson stars in some of these nonlinear sigma models inspired by string theory. Besides their connection with fundamental physics, these models are also interesting from a phenomenological point of view, as they can be regarded as nonlinear completions of the standard model \req{eq:SM}. While for small field values the nonlinear models will reduce to \req{eq:SM}, significant differences may arise for the most massive and compact boson stars, where the nonlinear effects kick in. Thus, they provide an interesting variation in the zoo of boson star solutions.

Quite surprisingly, these models have barely been studied in the boson star literature, and the few existing analyses \cite{Verbin:2007fa,Herdeiro:2018djx} --- both focusing on $\mathrm{O}(3)$ models --- have not yet provided a thorough exploration of the space of theories and solutions. 
Here we provide a detailed analysis of the effect of the curvature of the nonlinear sigma model on boson stars and show that it dramatically affects their masses and compactnesses. Thus, the sigma model metric is as relevant (or more) as the potential in determining the properties of boson stars. 

The paper is organized as follows. In section~\ref{sec:nonlinear} we review the nonlinear sigma models that will be the focus of our study. In section~\ref{sec:BS} we couple these models to gravity and study their spherically symmetric boson star solutions. We analyze some of their properties focusing on the dependence of their mass and compactness on the curvature of the sigma model. 
We discuss the relation between the energy scale of the nonlinear sigma model and the mass of boson stars in \ref{sec:energy}. Finally, we conclude in \ref{sec:discussion} by discussing our results as well as future directions.

\section{Nonlinear sigma models}\label{sec:nonlinear}
In this section we study the two-dimensional nonlinear sigma models (\textit{i.e.}, with two scalars) that are of potential interest for the construction of boson stars. We start discussing a particularly interesting model before presenting the general case.

\subsection{Axion-dilaton model}

The axion-dilaton system\footnote{One should not confuse the (stringy) axion-dilaton model with the (QCD) axion model of  \cite{Guerra:2019srj} --- these are totally unrelated theories. We also remark that, although the axion-dilaton system has sometimes been mentioned in the boson star literature \cite{Mielke:2001td}, a proper analysis of its solutions, taking into account its nonlinear character, has not been carried out.} is ubiquitous in the context of supergravity and string theory \cite{Ortin:2015hya}. The kinetic term of this model takes the form 
\begin{equation}\label{eq:axidilaton1}
\mathcal{L}_{K}=-\frac{1}{2}(\partial\phi)^2-\frac{1}{2}e^{2\gamma\phi}(\partial a)^2\, ,
\end{equation}
where $\phi$ is the dilaton, $a$ is the axion and $\gamma$ is a parameter that determines the strength of the coupling between both scalars. A very typical value in string theory scenarios is $\gamma=1$ (in Planck units), but other values are possible.  

The scalar manifold is the hyperbolic space and as such the kinetic term exhibits $\mathrm{SL}(2,\mathbb{R})$ symmetry. In order to make it manifest, it is convenient to work with the complex scalar field
\begin{equation}\label{taudef}
\tau=\gamma a+i e^{-\gamma\phi}\, ,
\end{equation}
in terms of which the Lagrangian reads
\begin{equation}
\mathcal{L}_{K}=-\frac{\partial_{\mu}\tau\partial^{\mu}\bar\tau}{2\gamma^2\text{Im}(\tau)^2}\, .
\end{equation}
Then, one can check that $\mathcal{L}_{K}$ is invariant under the $\mathrm{SL}(2,\mathbb{R})$  group in the form of M\"obius transformations
\begin{equation}
\tau'=\frac{c_1 \tau+c_2}{c_3 \tau+c_4}\, ,\qquad  \text{where} \, \,\,c_1c_4-c_2c_3=1\, ,
\end{equation}
with $c_i \in \mathbb{R}$. 

To complete our model, we need to provide a mass term for the scalars. This must preserve some of the symmetries of the kinetic term if the theory is to possess boson star solutions. It is not possible to have a potential that preserves the whole $\mathrm{SL}(2,\mathbb{R})$ group, but however it is enough to focus on the $\mathrm{U}(1)$ subgroup given by the transformation
\begin{equation}\label{eq:U1sym}
\tau'=\frac{\tau \cos(\alpha/2)+\sin(\alpha/2)}{\cos(\alpha/2)-\tau \sin(\alpha/2)}\, ,\quad  \alpha\in [0,2\pi)\, .
\end{equation}
Then, we look for a potential $U(\tau, \bar\tau)$ that is invariant under such transformation. In particular, invariance of $U$ under the infinitesimal transformation $\delta\tau=\frac{1}{2}(1+\tau^2)$ implies that
\begin{equation}
\delta U=\frac{1}{2}(1+\tau^2)\partial_{\tau}U+\frac{1}{2}(1+\bar\tau^2)\partial_{\bar\tau}U=0
\end{equation}
Solving this equation we find the the potential must have the form $U=U(\mathcal{T}^2)$, where $\mathcal{T}^2$ is the basic invariant
\begin{align}\notag
\mathcal{T}^2&=\frac{1}{\gamma^2}\left(\frac{1+|\tau|^2}{\text{Im}(\tau)}-2\right)\\
&=\frac{4}{\gamma^2}\sinh^2\left(\frac{\gamma\phi}{2}\right)+a^2 e^{\gamma\phi}\, .
\end{align}

Summarizing, our theory with a massive axidilaton with $\mathrm{U}(1)$ symmetry is in general given by
\begin{equation}\label{eq:axidilaton}
\mathcal{L}=-\frac{\partial_{\mu}\tau\partial^{\mu}\bar\tau}{2\gamma^2\text{Im}(\tau)^2}-U\left(\mathcal{T}^2\right)\, ,
\end{equation}
and it only remains to choose a specific potential. A natural choice would be a polynomial in $\mathcal{T}^2$, the simplest choice being just $U(\mathcal{T}^2)=\frac{1}{2}\mu^2\mathcal{T}^2$, where $\mu$ is the mass of the field.  As we explain below, this is the analogous of a non-interactive potential in the case of a curved sigma model --- see Footnote \ref{foot}.

\subsection{Maximally symmetric models}
The axion-dilaton model is a maximally symmetric sigma model in the sense that its target manifold --- the hyperbolic space --- is maximally symmetric.\footnote{Of course, the potential always breaks this symmetry down to $\mathrm{U}(1)$.}
We can also consider its positive-curvature version, the $\mathrm{O}(3)$ model, where the target manifold is the sphere. We can write the two theories in a unified form as follows. Starting from the axion-dilaton model \req{eq:axidilaton} we perform the change of variables
\begin{equation}
\Phi=\frac{2}{\gamma}\frac{1+i\tau}{1-i\tau}\, ,
\end{equation} 
in terms of which the Lagrangian reads
\begin{equation}\label{eq:axidilatonPhi}
\mathcal{L}=-\frac{\partial_{\mu}\Phi\partial^{\mu}\bar{\Phi}}{2\left(1-\frac{\gamma^2}{4}|\Phi|^2\right)^2}-U\left(\frac{|\Phi|^2}{\left(1-\frac{\gamma^2}{4}|\Phi|^2\right)}\right)\, ,
\end{equation}
where we used that 
\begin{equation}
\mathcal{T}^2=\frac{|\Phi|^2}{\left(1-\frac{\gamma^2}{4}|\Phi|^2\right)}\, .
\end{equation}
When expressed in this form we observe that the theory only depends on $\gamma^2$, allowing us to extend it to $\gamma^2<0$. In fact, while for $\gamma^2>0$ the scalar manifold is the hyperbolic space --- hence the $\mathrm{SL}(2,\mathbb{R})$ symmetry ---  for $\gamma^2<0$ it corresponds to the 2-sphere, so we get an $\mathrm{O}(3)$ sigma model. For $\gamma=0$ it reduces to the linear sigma model \req{eq:SM}. Indeed, these are the three possible maximally symmetric sigma models, where $\gamma^2$ is nothing but (minus) the curvature of the scalar manifold.

\subsection{General case}
Another interesting aspect about \req{eq:axidilatonPhi} is that now the $\mathrm{U}(1)$ symmetry is manifest, as it acts on $\Phi$ in the usual way 
\begin{equation}
\Phi\rightarrow e^{i\alpha} \Phi\, .
\end{equation}
In fact, in Appendix \ref{app:sigmamodel} we show that every two-dimensional nonlinear sigma model with a $\mathrm{U}(1)$ holomorphic isometry can be written in the ``canonical'' form 
\begin{equation}\label{eq:canonical}
\mathcal{L}^{\mathrm{U}(1)}_{\Phi}=-K\left(|\Phi|^2\right)\partial_{\mu}\Phi\partial^{\mu}\bar{\Phi}-V\left(|\Phi|^2\right)\, ,
\end{equation}
with a certain kinetic function $K\left(|\Phi|^2\right)$ and potential $V\left(|\Phi|^2\right)$. The Lagrangian \req{eq:axidilatonPhi} is in this sense the canonical form of the maximally symmetric theories. 

In closing this section, let us take note of the conserved current associated to the unbroken $\mathrm{U}(1)$ symmetry of these models, 
\begin{equation}
J_{\mu}=i K\left(|\Phi|^2\right)\left(\bar{\Phi}\partial_{\mu}\Phi-\Phi\partial_{\mu}\bar{\Phi}\right)\, .
\end{equation}

\section{Boson stars}\label{sec:BS}
\subsection{Theory}
Let us now couple the general $\mathrm{U}(1)$ invariant model \req{eq:canonical} to gravity,
\begin{equation}\label{theory}
S=\frac{1}{16\pi G}\int d^4x\sqrt{-g}\left[R+\mathcal{L}^{\mathrm{U}(1)}_{\Phi}\right]\, .
\end{equation}
We remark that the scalar field is expressed in Planck units, hence the common factor of $(16\pi G)^{-1}$. We will come back to this when discussing the energy scale of the sigma model in Section~\ref{sec:energy}.

The equations of motion of this theory read
\begin{align}\label{eq:einstein}
G_{\mu\nu}=&K\left(\partial_{(\mu}\Phi\partial_{\nu)}\bar\Phi-\frac{1}{2}g_{\mu\nu}\partial_{\alpha}\Phi\partial^{\alpha}\bar{\Phi}\right)-\frac{1}{2}Vg_{\mu\nu}\, ,\\
0=&K\nabla^2\Phi+K' \bar\Phi \left(\partial\Phi\right)^2-\Phi V'\, ,
\label{eq:scalar}
\end{align}
where $G_{\mu\nu}$ is the Einstein tensor and $K'=dK/d|\Phi|^2$, $V'=dV/d|\Phi|^2$.  We focus here on the family of models \req{eq:axidilatonPhi}, so we choose 
\begin{align}
K\left(|\Phi|^2\right)&=\frac{1}{2\left(1-\frac{\gamma^2}{4}|\Phi|^2\right)^2}\, .
\end{align}
In addition, we use the following potential
\begin{align}\label{eq:potentialPHI}
V\left(|\Phi|^2\right)=\frac{\mu^2|\Phi|^2}{2\left(1-\frac{\gamma^2}{4}|\Phi|^2\right)}\, ,
\end{align}
where $\mu$ represents the mass of the bosons. 
This is analogous to a non-interactive mass term in the case of a curved sigma model,\footnote{\label{foot}This is the only potential for which plane waves $\Phi=A e^{-i k_{\mu}x^{\mu}}$, with $k^2=-\mu^2$, remain exact solutions. Hence this is the closest one can get to free fields in the case of a curved sigma model.} while interaction terms would correspond to higher powers of the same combination. We do not include those since our goal is to study the effect of the curvature of the sigma model on boson stars, rather than the effect of the potential. Thus, we stick to the minimally interacting potential  \req{eq:potentialPHI}.
 
\subsection{Boson star ansatz}
We look for spherically symmetric solutions of the form
\begin{align}\label{eq:metric}
ds^2=&-f(r)dt^2+\frac{dr^2}{g(r)}+r^2d\Omega_{(2)}^2\, ,\\
\Phi=&\Phi_0(r)e^{-i\omega t}\, ,
\end{align}
where $\Phi_0(r)$ is real. The harmonic time dependence of $\Phi$ ensures that the stress-energy tensor is stationary even if the scalar field is not. Now, instead of working with $\Phi_0(r)$, we find it interesting to work  with the variable 
\begin{equation}
\phi(r)=\begin{cases}
\frac{2}{\gamma}\operatorname{argtanh}\left(\frac{\gamma \Phi_0(r)}{2}\right) &\quad \text{if}\,\, \gamma^2>0\\
\Phi_0(r) &\quad \text{if}\,\, \gamma=0\\
\frac{2}{|\gamma|}\arctan\left(\frac{|\gamma| \Phi_0(r)}{2}\right) &\quad \text{if}\,\, \gamma^2<0\, .
\end{cases}
\end{equation}
This can be identified with the original dilaton in \req{eq:axidilaton1} in the case of $\gamma^2>0$. 
With this, we find that the $tt$ and $rr$ components of Einstein equations \req{eq:einstein} read
\begin{align}\label{eq:geq}
r g'+g-1&=-2r^2\rho\\
g\left(\frac{r f'}{f}+1\right)-1&=2r^2p \, ,
\label{eq:feq}
\end{align}
while the scalar equation \req{eq:scalar} yields
\begin{equation}\label{eq:phieq}
\begin{aligned}
\phi ''+\frac{1}{2} \phi ' \left(\frac{f'}{f}+\frac{g'}{g}+\frac{4}{r}\right)+\frac{\omega^2 }{f g} \mathcal{S}_{2\gamma}(\phi)=\frac{\mu^2}{g}\mathcal{S}_{\gamma}(\phi)\, .
\end{aligned}
\end{equation}
Here the density $\rho=f^{-1}T_{tt}$ and radial pressure $p=gT_{rr}$  are given by
\begin{align}
\rho=&+\frac{\mu^2}{8}\mathcal{S}^2_{\gamma/2}(\phi)+\frac{g}{8}\phi'^2+\frac{\omega^2}{8f}\mathcal{S}^2_{\gamma}(\phi)\, ,\\
p=&-\frac{\mu^2}{8}\mathcal{S}^2_{\gamma/2}(\phi)+\frac{g}{8}\phi'^2+\frac{\omega^2}{8f }\mathcal{S}^2_{\gamma}(\phi)\, ,
\end{align}
and we have introduced the notation
\begin{equation}
\mathcal{S}_{\gamma}(\phi):=\begin{cases}
\frac{1}{\gamma}\sinh(\gamma \phi) &\quad \text{if}\,\, \gamma^2>0\\
\phi &\quad \text{if}\,\, \gamma=0\\
\frac{1}{|\gamma|}\sin(|\gamma| \phi) &\quad \text{if}\,\, \gamma^2<0\, .
\end{cases}
\end{equation}
We note that the appearance of the function $\mathcal{S}_{\gamma}(\phi)$ is the key effect of the nonlinear sigma models.  We also remark that in the case $\gamma^2<0$ the variable $\phi(r)$ is compact, as it lies in the interval $-\pi |\gamma|< \phi< \pi |\gamma|$.

Boson stars are horizonless solutions that are regular at the origin. To ensure smoothness, we must impose $g(0)=1$, and the Taylor expansion of the functions $f$, $g$ and $\phi$ around $r=0$ must only contain even powers of $r$, 
\begin{equation}\label{eq:BC}
\begin{aligned}
f(r)&=f_0+\sum_{n=1}^{\infty}f_{2n}r^{2n}\, ,\\
g(r)&=1+\sum_{n=1}^{\infty}g_{2n}r^{2n}\, ,\\
\phi(r)&=\phi_0+\sum_{n=1}^{\infty}\phi_{2n}r^{2n}\, .
\end{aligned}
\end{equation}
The whole series is determined by $f_0$ (which must be taken positive) and the central value of the scalar field $\phi_0$. For instance, the $\mathcal{O}(r^2)$ coefficients are given by
\begin{align}
f_{2}&=\frac{1}{6} \omega ^2 \mathcal{S}_{\gamma }^2\left(\phi _0\right)-\frac{1}{12} f_0 \mu ^2 \mathcal{S}_{\gamma/2}^2\left(\phi _0\right)\, ,\\
g_{2}&=-\frac{\left(f_0 \mu ^2 \mathcal{S}_{\gamma/2}^2\left(\phi _0\right)+\omega ^2 \mathcal{S}_{\gamma}^2\left(\phi _0\right)\right)}{12 f_0}\, ,\\
\phi_{2}&=\frac{1}{6} \left(\mu ^2 S_{\gamma }\left(\phi _0\right)-\frac{\omega ^2 S_{2 \gamma }\left(\phi _0\right)}{f_0}\right)\, ,
\end{align}
and one can similarly find the subsequent terms. 


On the other hand, at infinity we require that the solution be asymptotically flat. This will happen if,
\begin{equation}
\begin{aligned}
f(r\rightarrow \infty )=g(r\rightarrow \infty )&=1\, ,\\
\phi(r\rightarrow\infty)&=0\, .
\end{aligned}
\end{equation}
The condition on $f$ fixes the value of $f_0$, while the condition on $g$ is automatically satisfied taking into account Eq.~\req{eq:geq} and that $r^2\rho\rightarrow 0$ at infinity. On the other hand, $\phi$ has the familiar Yukawa-type behavior near infinity,
\begin{equation}
\phi(r)\sim \frac{\phi_{+}}{r}e^{r\sqrt{\mu^2-\omega^2}}+\frac{\phi_{-}}{r}e^{-r\sqrt{\mu^2-\omega^2}}
\end{equation}
For $\omega^2<\mu^2$ we have a exponentially growing mode and a exponentially decaying one, and hence asymptotic flatness is only achieved for $\phi_{+}=0$. However, for a given choice of $\phi_0$ and $\omega$ the solution defined by the regularity conditions \req{eq:BC} typically contains both modes and hence is not asymptotically flat.\footnote{In practice, those solutions cannot be extended to infinity as they typically develop a singularity at some finite $r$.} Thus, the asymptotic flatness condition provides an eigenvalue problem that only has solutions for discrete values of $\omega$ that depend on $\phi_0$. These are the bound states representing boson stars. These solutions can be labeled by an integer $n$ that characterizes the number of nodes of the scalar field. Here we will focus on the ground state solution $n=1$.
Finally, for $\omega^2>\mu^2$ the scalar field radiates and we do not have confined solutions. 

\subsection{Numerical strategy}
In order to solve the differential equations \req{eq:feq}, \req{eq:geq}, \req{eq:phieq} we can always work in natural units with $\mu=1$ by performing $r\rightarrow r/\mu$, $\omega\rightarrow \omega \mu$. In addition, it is also interesting to set $f_0=1$, which is equivalent to performing the rescalings
\begin{equation}
t=\tilde t /\sqrt{f_0}\, ,\quad f=f_0 \tilde{f}\, ,\quad \omega=\sqrt{f_0}\tilde{\omega}\, .
\end{equation}
The value of $f_0$ can then be retrieved once a solution is found by using that $\lim_{r\rightarrow\infty} \tilde f(r)=1/f_0$

\begin{figure}[t!]
	\centering
	\includegraphics[width=0.45\textwidth]{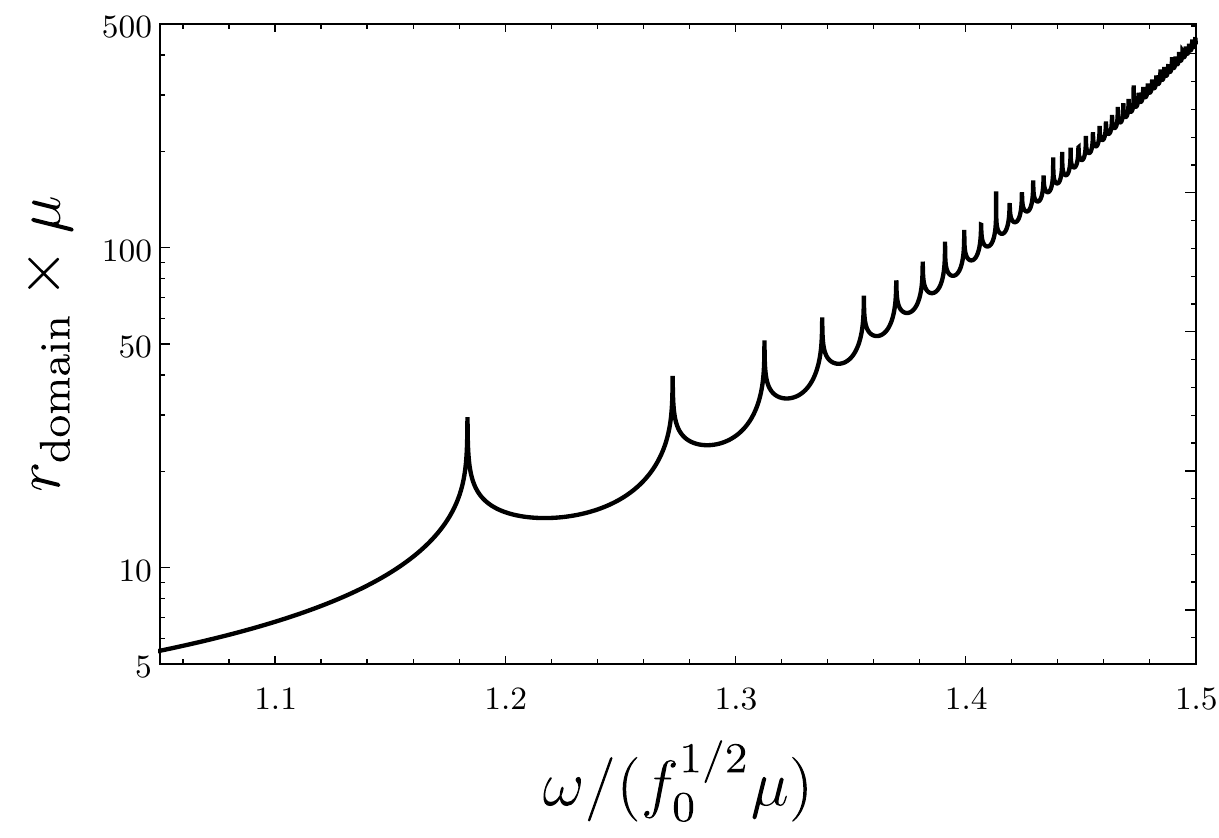}
	\caption{Radius of existence of the solutions with $\phi_0=0.5$, $\gamma=1$ as a function of $\omega$. Boson stars are the solutions whose domain extends to infinity and they correspond to the peaks shown in this curve, with the ground-state boson star being the leftmost peak.}
	\label{fig:domainr}			
\end{figure}

We solve \req{eq:feq}, \req{eq:geq}, \req{eq:phieq} numerically using the expansion \req{eq:BC} at order $r^4$ to set our initial conditions near $r=0$ (we used $r_{\rm in }=10^{-3}$). Generically, for a given  $\omega$, the solution develops a singularity at some finite $r=r_{\rm domain}$ and hence it does not extend to infinity. However, when $\omega$ approaches some special values, the domain of existence grows indefinitely, pointing the existence of asymptotically flat solutions. We show this domain of existence in Fig.~\ref{fig:domainr} for $\gamma=1$ and $\phi_0=1/2$. The peaks in the curve correspond to the boson star solutions, with the leftmost peak being the ground state $n=1$. 
In order to find $\omega$ with higher accuracy, we then employed a bisection-type method that searches the position of these peaks. 
As we can observe, the peaks are extremely sharp, and in order to obtain accurate enough solutions one needs to find $\omega$ with many digits of precision. The problem becomes stiffer for larger  values of $|\gamma|$ and $\phi_0$ (corresponding to the most compact solutions), which makes obtaining those solutions particularly challenging. We provide additional details on our numerical approach in Appendix~\ref{app:numerics}.

\subsection{Mass, charge and radius} 
Once a numerical solution has been found, we can obtain some of its properties. 
Let us introduce the mass function $m(r)$ by
\begin{equation}
g(r)=1-\frac{2G m(r)}{r}\, .
\end{equation}
The total ADM mass of the solutions is given by the asymptotic value 
\begin{equation}
M=\lim_{r\rightarrow\infty} m(r)\, .
\end{equation}
On the other hand, boson stars do not have a sharp surface, but it is usual to define their radius $R$ as the value of $r$ that contains 99\% of the mass. Thus, $R$ is implicitly defined by
\begin{equation}
\frac{m(R)}{M}=0.99\, .
\end{equation}

\begin{figure*}[t!]
	\centering
	\includegraphics[width=0.47\textwidth]{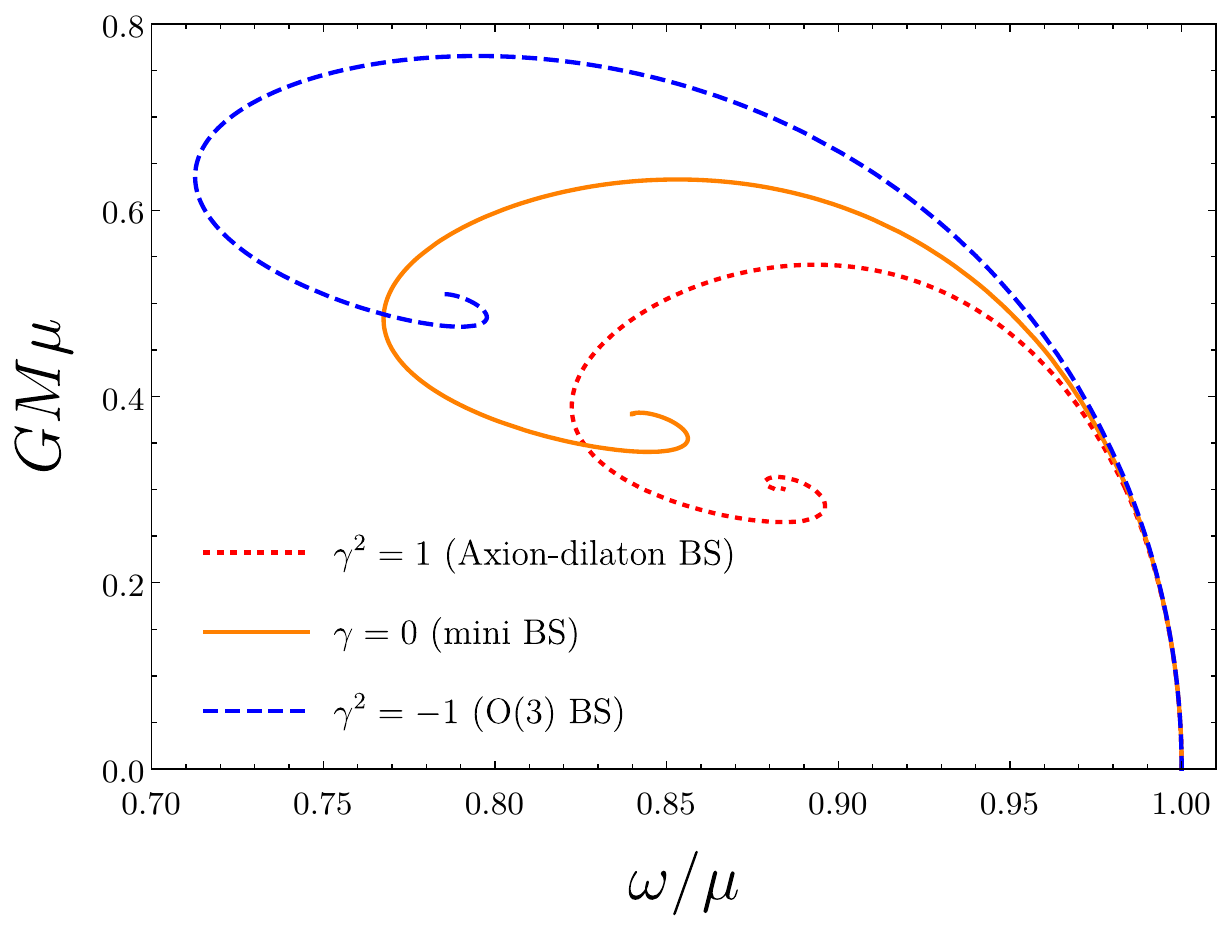}\hskip0.5cm
	\includegraphics[width=0.47\textwidth]{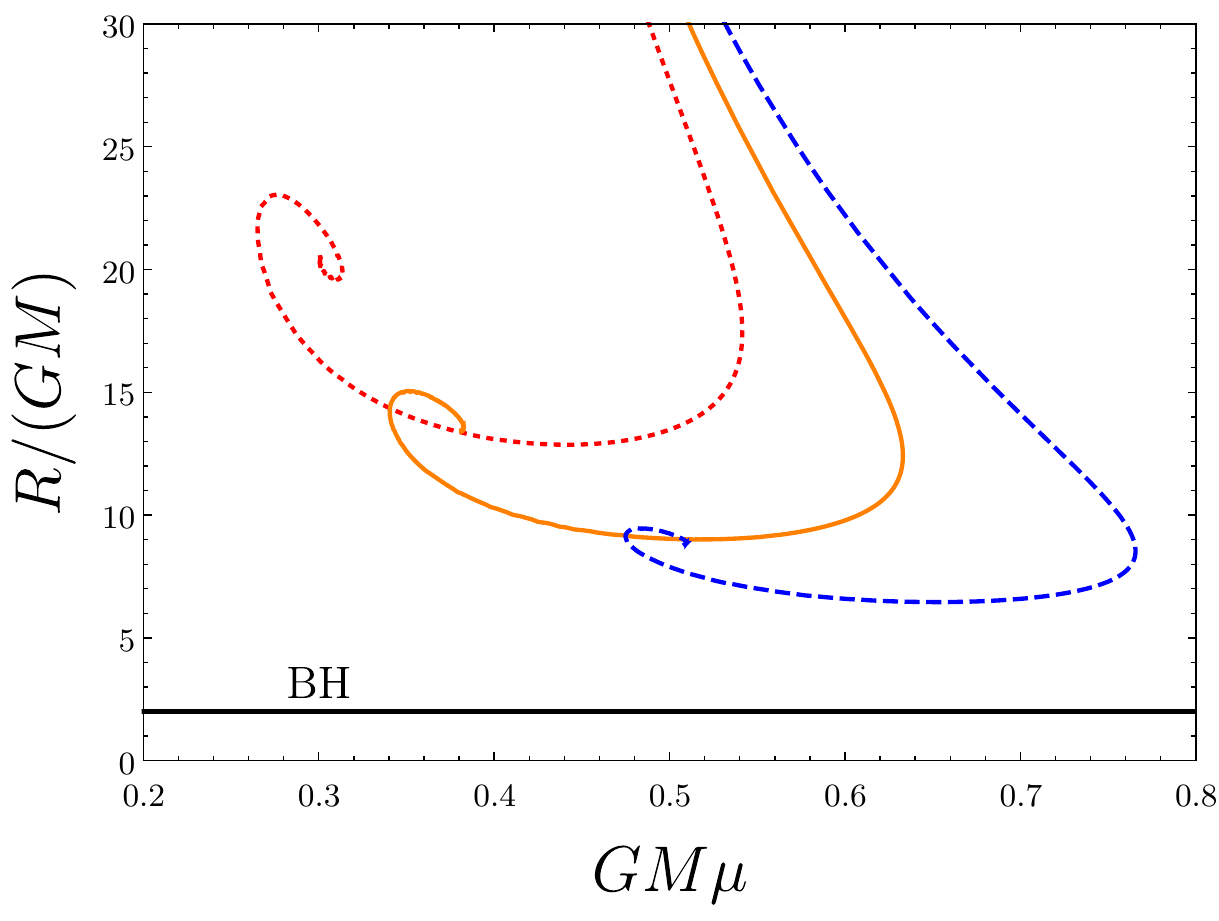}
	\caption{Left: existence curve $(\omega, M)$ for ground-state boson stars with different values of $\gamma$. Right: compactness $R/(GM)$ as a function of the mass for the same families of boson stars.}
	\label{fig:existencecurve}			
\end{figure*}  
\begin{figure*}[t!]
			\centering
			\includegraphics[width=0.49\textwidth]{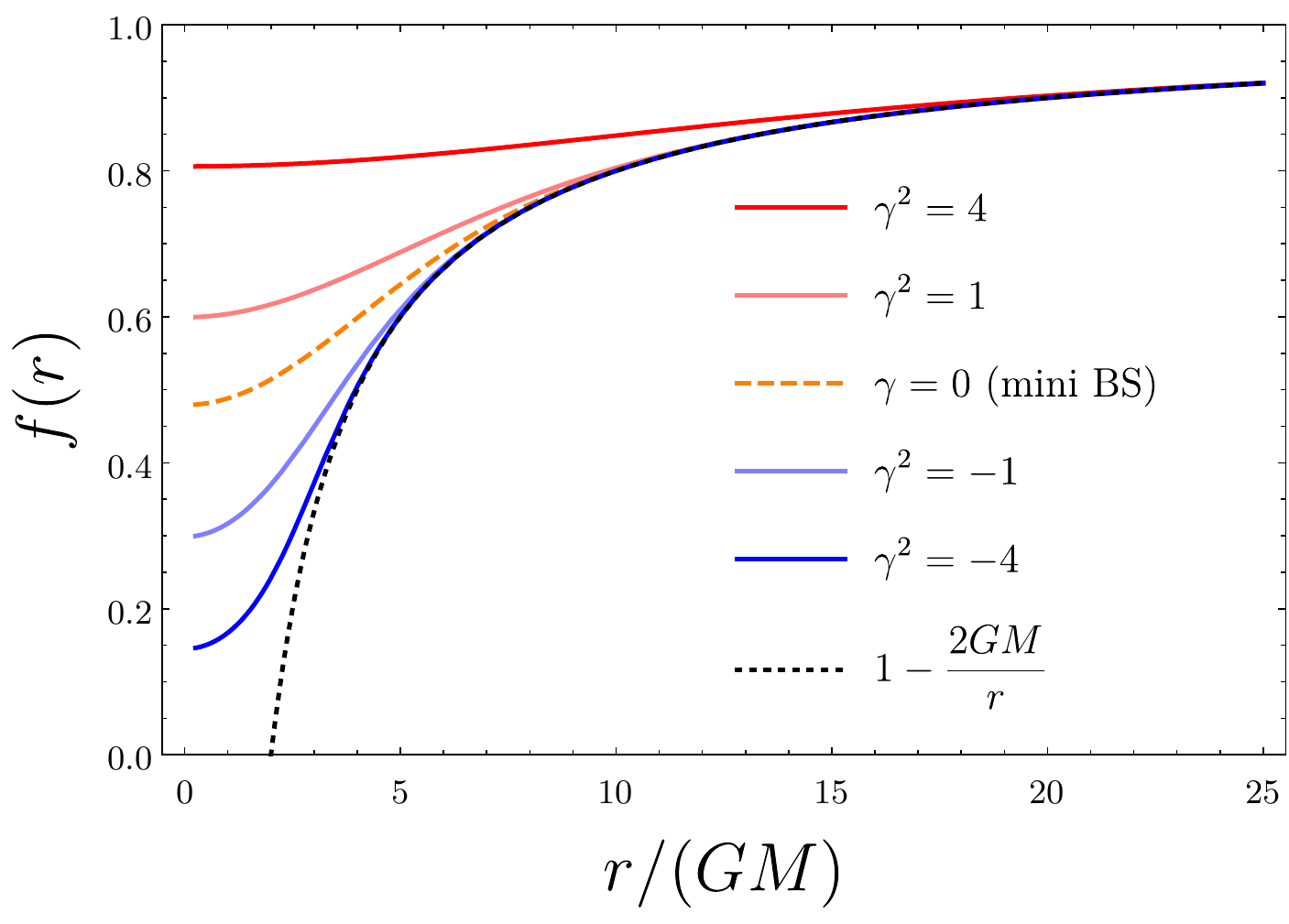}
			\includegraphics[width=0.49\textwidth]{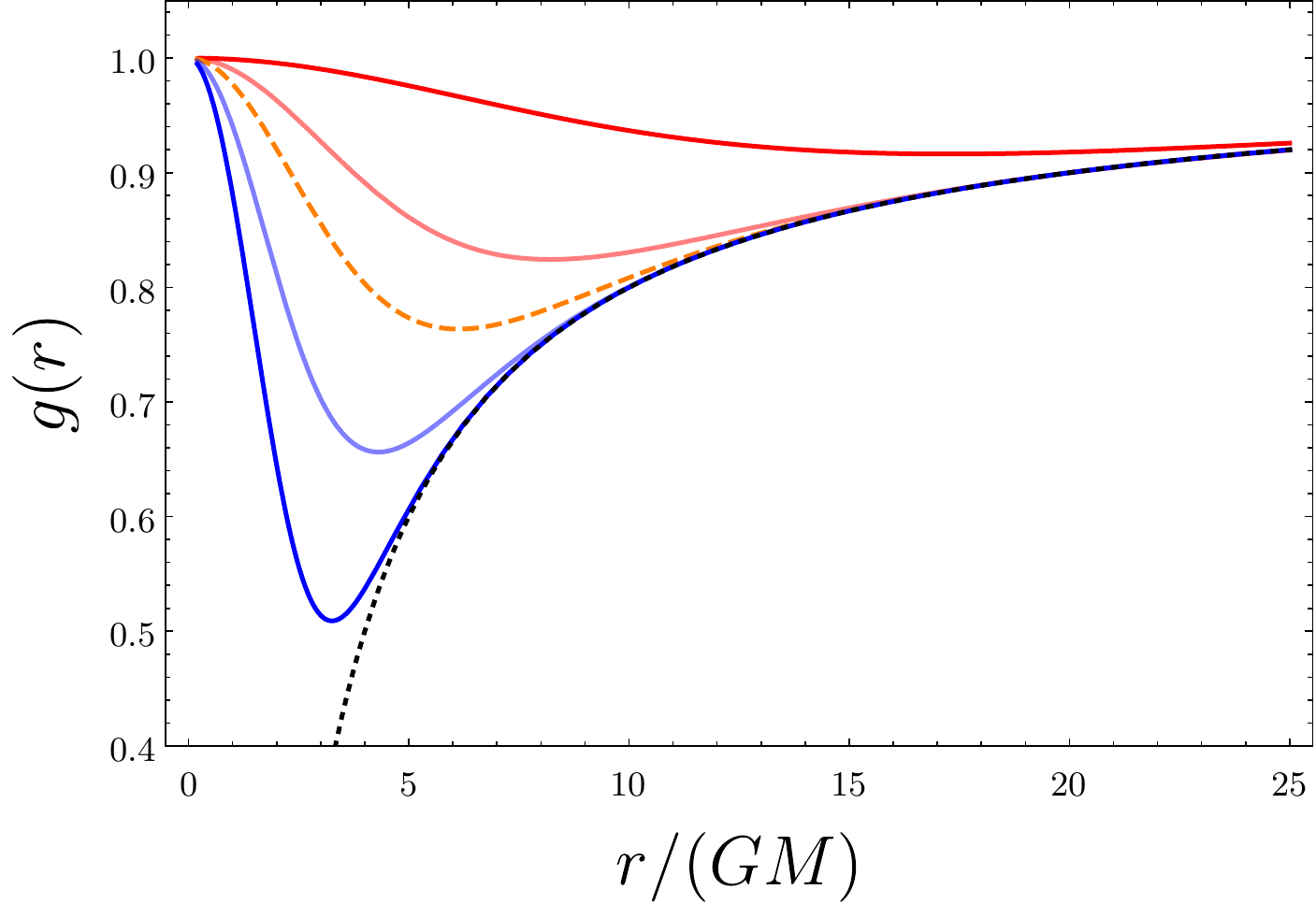}
			\caption{Metric functions $f(r)$ and $g(r)$ for the maximum mass solutions corresponding to several values of $\gamma$. For comparison, we also show the Schwarzschild functions $f(r)=g(r)=1-2GM/r$ (black dotted line).}
			\label{fig:metricfunctions}			
\end{figure*}

We shall define the compactness as the ratio $R/(GM)$, which takes a minimum value of $2$ for a Schwarzschild black hole.\footnote{We take note of the confusion that a \emph{greater} compactness means a \emph{smaller} $R/(GM)$. Other authors refer to the inverse of this quantity as the compactness in order to avoid this issue. However, we prefer to report the values of $R/(GM)$ because it is a more intuitive number.} 
Finally, the conserved $\mathrm{U}(1)$ charge reads
\begin{equation}
Q=\frac{\omega}{4G}\int_{0}^{\infty}dr\frac{r^2\mathcal{S}_{\gamma}^2(\phi)}{\sqrt{f(r)g(r)}}
\end{equation}

\subsection{Results}
In the left panel of Fig.~\ref{fig:existencecurve} we show the existence curve $M$ versus $\omega$ for ground-state mini boson stars, axion-dilaton stars with $\gamma^2=1$ and and  $\mathrm{O}(3)$ stars with $\gamma^2=-1$. These curves are generated by varying the value of $\phi_0$ from $0$ (corresponding to the deconfinement limit $\omega\rightarrow \mu$) to a maximum value that depends on the case. We see that the existence curves in all cases have the same qualitative profile as for mini boson stars, with the characteristic spiral shape. We observe quantitative differences though: the axion-dilaton solutions reach smaller masses and exist for a smaller range of frequencies than the mini boson stars. The opposite effect is observed for the $\mathrm{O}(3)$ model: these solutions reach higher masses and exists for a bigger range of frequencies. These effects become more drastic as we increase $\gamma^2$ (in absolute value). On the other hand, in the deconfinement limit $\omega\rightarrow \mu$ (so $\phi_0\rightarrow 0$) the three curves converge. This is expected since for small values of $\phi$ the curvature of the scalar manifold becomes irrelevant and all the models reduce to the mini boson star one.  
 
The right panel of Fig.~\ref{fig:existencecurve} shows the compactness $R/(GM)$ for these families of boson stars. Here we can see that $\mathrm{O}(3)$ stars are not only more massive, but also more compact. Therefore, it appears that a positively curved sigma model leads to more extreme boson stars, while negative curvature tends to make them more diffuse.

The maximum mass solution in these diagrams plays an important role. Generally, one expects only the branch of solutions that connects the maximum mass solution to the $\omega\rightarrow \mu$ solution to be stable. Those that lie beyond the maximum mass solution are in principle unstable.\footnote{However, the non-linear character of the theories at hand could change this behavior, so it would be interesting to perform an analysis of the stability of the solutions.}
Within the naively stable branch, the maximum mass solutions are also the most compact ones, which makes them particularly interesting. We thus focus on these solutions in what follows. 

We show the profile of the metric functions $f(r)$ and $g(r)$ for the maximum mass solutions for several values of $\gamma$ in Fig.~\ref{fig:metricfunctions}. Here we show the radius in units of the corresponding mass $M$ for an easier comparison, but we recall that for a fixed $\mu$ all of these solutions have different masses. This figure clearly shows that the $\mathrm{O}(3)$ boson stars are more compact and produce a deeper gravitational well than their axion-dilaton counterparts. In fact, the effect is very drastic. Already for $\gamma=2$ the axion-dilaton stars become so disperse that they can probably be well described by the Newtonian approximation.  The other side of the coin is that $\mathrm{O}(3)$ stars resemble more and more a black hole as we increase $|\gamma|$. 

\begin{figure}[t!]
			\centering
			\includegraphics[width=0.49\textwidth]{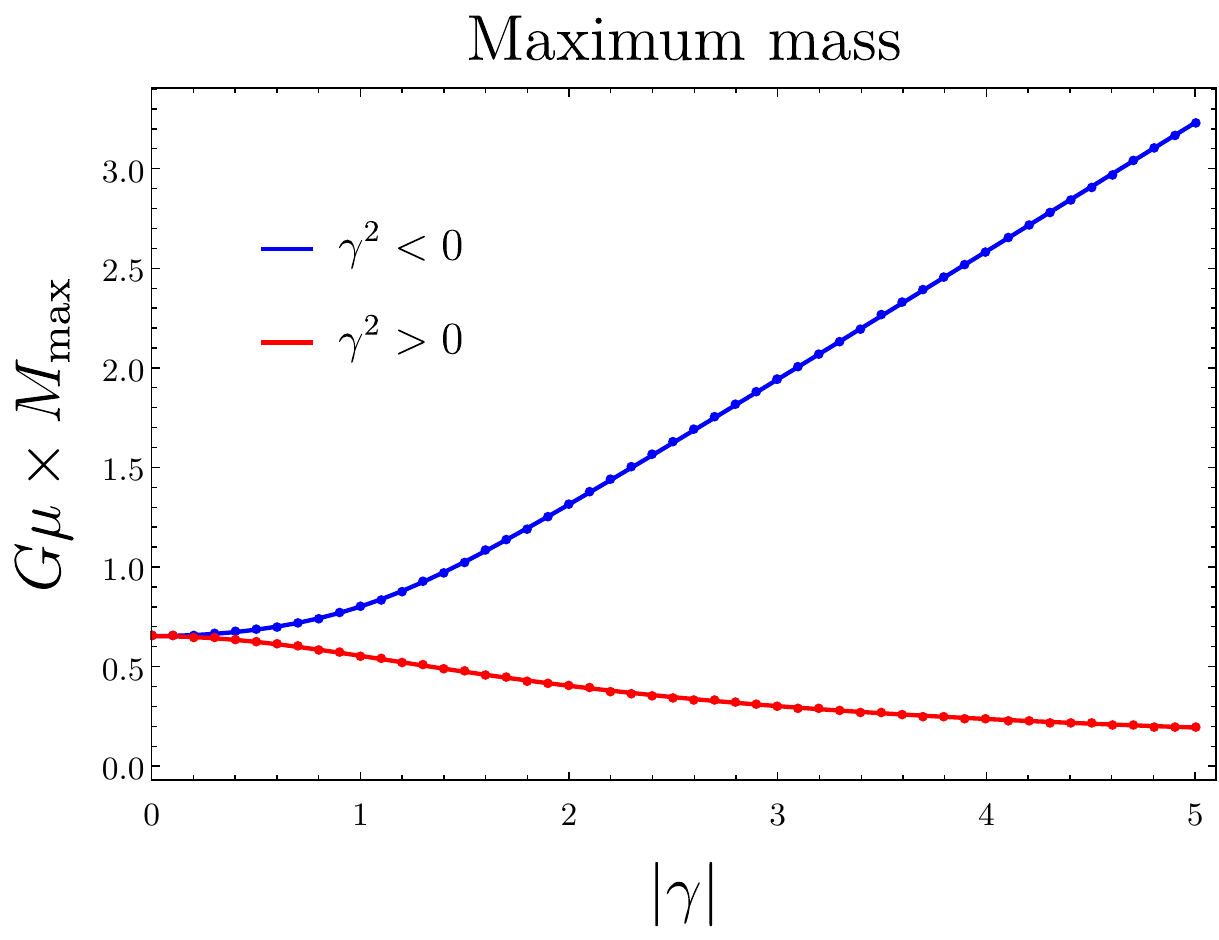}
			\caption{Maximum mass of boson stars as a function of $|\gamma|$ for the $\mathrm{SL}(2,\mathbb{R})$ $(\gamma^2>0)$ and $\mathrm{O}(3)$ $(\gamma^2<0)$ models. When $|\gamma|$ is large, the maximum mass of $\mathrm{O}(3)$ stars is proportional to $|\gamma|$, while in the case of $\mathrm{SL}(2,\mathbb{R})$ stars we find that it is inversely proportional. }
			\label{fig:maxmass}			
\end{figure} 

\begin{figure*}[t!]
			\centering
			\includegraphics[width=0.48\textwidth]{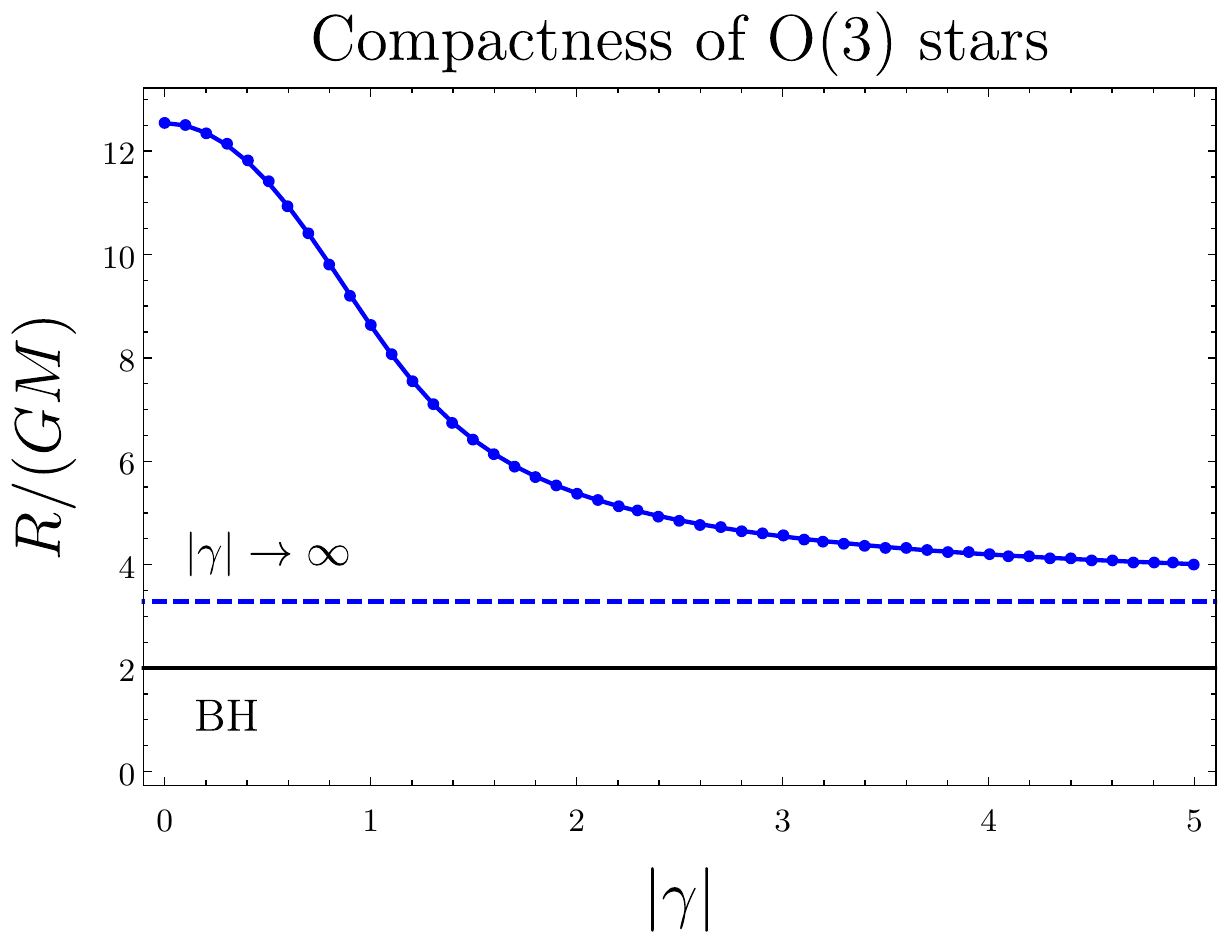}
			\includegraphics[width=0.49\textwidth]{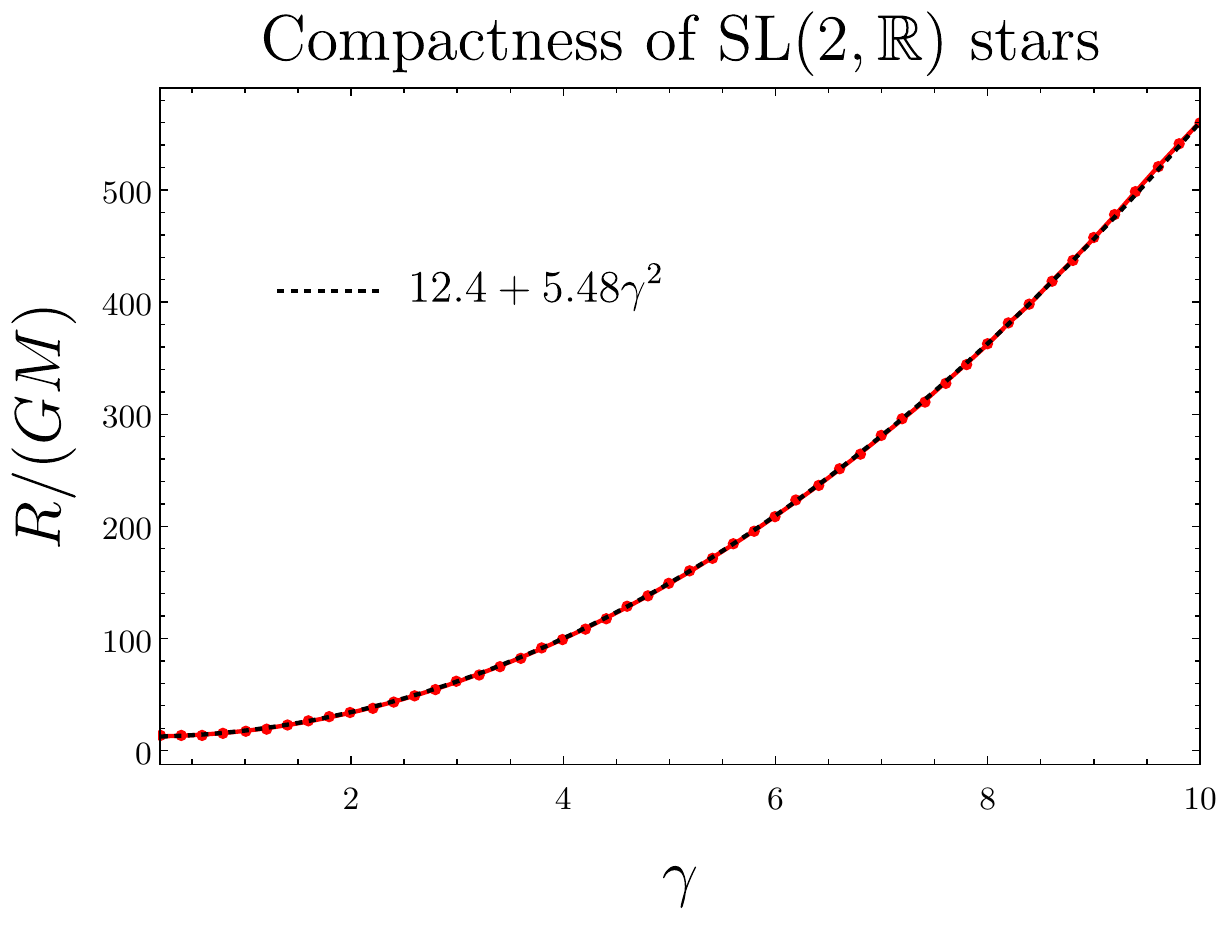}
			\caption{Compactness of  the maximum mass $\mathrm{O}(3)$ and $\mathrm{SL}(2,\mathbb{R})$ boson stars as a function of $|\gamma|$. The dashed line in the left plot represents the limit of the compactness of the $\mathrm{O}(3)$ solutions when $|\gamma|\rightarrow \infty$, approximately $R/(GM)\approx 3.29$. 	In the right plot, along with our numerical results for the compactness of $\mathrm{SL}(2,\mathbb{R})$ stars, we show a quadratic fit that seems to capture perfectly the behavior of $R/(GM)$ for all values of $\gamma$.}
			\label{fig:compactness}			
\end{figure*}

It is then interesting to look at the dependence on $\gamma$ of these solutions. 
In Fig.~\ref{fig:maxmass}, we show the maximum mass as a function of $|\gamma|$, for $\gamma^2>0$ and $\gamma^2<0$, while in Fig.~\ref{fig:compactness} we show the compactness of these solutions.

Let us take a look first at the $\mathrm{O}(3)$ solutions. The most obvious thing we observe in Fig.~\ref{fig:maxmass} is that the relation between $M_{\rm max}$ and $|\gamma|$ quickly becomes linear for large $|\gamma|$. In fact, by performing a linear fit to the numerical data we find that
\begin{equation}\label{mmaxo3}
M_{\rm max}^{\rm \small O(3)}\approx 0.565\frac{|\gamma|}{G \mu}
\end{equation}
when $|\gamma|\gg1$. The compactness of these solutions --- as illustrated in the left plot of Fig.~\ref{fig:compactness} --- also increases with $|\gamma|$, with a maximum compactness that would be reached for $|\gamma|\rightarrow\infty$. 
It is difficult to estimate this value since it depends on the asymptotic behavior of $R/(GM)$ as a function of $|\gamma|$, which we do not know a priori. Our data strongly suggests a power-law decay and, by trying different fits, it favors an expansion of the form $R/(GM)=C_0+C_1|\gamma|^{-1}+C_3 |\gamma|^{-3}+\mathcal{O}(|\gamma|^{-5})$. By assuming this, we obtain the following fit
\begin{equation}
\frac{R}{GM}\bigg|_{M_{\rm max}^{\rm \small O(3)}}\approx 3.29+\frac{3.2}{|\gamma|} +\mathcal{O}\left(|\gamma|^{-3}\right)\quad \text{for}\,\, |\gamma|\gg 1\, .
\end{equation}
Thus, for large $|\gamma|$ these stars are close to the compactness  $R\sim3GM$, which is usually the threshold for ultracompact objects --- those with light rings. While our numerical solutions up to $|\gamma|=5$ do not show light rings, we cannot at this point discard their existence for larger values of $|\gamma|$, for which the numerical computations become more involved.\footnote{Light rings do arise in boson stars in the unstable branch, as already happens in the case of mini boson stars \cite{Cunha:2017wao}. Our results show that in the case of $\mathrm{O(3)}$ stars, light rings arise closer to the maximum mass solution as we increase $|\gamma|$, but it is unclear at this point whether for larger $|\gamma|$ these solutions move into the stable branch.} 

On the other hand, the behaviour of axion-dilaton stars ($\gamma^2>0$) is radically opposite to their $\mathrm{O}(3)$ counterparts, with their maximum mass and compactness quickly decreasing with $\gamma$. Performing a fit of the form $M_{\rm max}=M_0+M_1/\gamma+M_2/\gamma^2+\ldots$, we find that the results are consistent with $M_0=0$, \textit{i.e.}, the mass tends to zero for large $\gamma$. Indeed, we find that 
\begin{equation}\label{mmaxsl2}
M_{\rm max}^{\rm \small SL(2,\mathbb{R})}\approx \frac{1.1}{G \mu \gamma}+\mathcal{O}\left(\gamma^{-2}\right)\quad \text{for}\,\, \gamma\gg 1\, .
\end{equation}
In addition,  the ratio $R/(GM)$ (right plot of Fig.~\ref{fig:compactness}) grows very fast for large values of $\gamma$. Our results indicate that it scales with $\gamma^2$ when $\gamma\gg 1$, and in fact, a simple quadratic polynomial fit captures our numerical values with very good accuracy in all the range of $\gamma$: 
\begin{equation}
\frac{R}{GM}\bigg|_{M_{\rm max}^{\rm \small SL(2,\mathbb{R})}}\approx 12.4+5.48\gamma^2\, .
\end{equation}
Thus, these solutions quickly acquire very low compactnesses compared to those of mini boson stars and therefore, their gravitational field is essentially Newtonian. 

Finally, we provide a few additional properties of the maximum mass solutions for a selection of values of $\gamma$ in Table~\ref{t1}.
It is worth emphasizing that in all the cases the charge of solutions is very similar to the mass --- they are roughly proportional across all values of $\gamma$. In addition, we observe that the frequency of the  $\mathrm{SL}(2,\mathbb{R})$ solutions approaches $\omega\rightarrow \mu$ when $\gamma$ is large, naturally corresponding to the decompactification limit. In the case of $\mathrm{O}(3)$ stars, the frequency tends to a non-zero constant.  In fact, our results suggest that $\omega\rightarrow \tfrac{2}{3}\mu$ for $|\gamma|\rightarrow\infty$.

\begin{table}[h!]
\centering
\begin{tabular}{|c|c|c|c|c|c|}
	\hline
	$\gamma$ & $\phi_0$ & $\omega/\mu$    & $M\times G \mu$  & $Q\times G \mu$ & $R/(GM)$ \\
	\hline\hline
	0 &0.5431& 0.8529 & 0.6330 & 0.6530 & 12.4\\\hline
	1 & 0.4078 & 0.8932 &0.5362 & 0.5538 &17.6\\
	2 & 0.2222 & 0.9433 &0.3947 &0.4034 &33.9\\
	5 &0.0523 & 0.9868 & 0.1914 & 0.1938 &149.0\\	
		\hline
	$i$ &0.6855 & 0.7949 & 0.7653 & 0.8012 & 8.54\\
	$2 i$ & 0.5983 & 0.7139 & 1.2065 & 1.3144 & 5.33\\
	$5 i$ & 0.2637 & 0.6771 & 2.8564 &3.2307 &  3.97\\
	\hline
\end{tabular}
\caption{Properties of the maximum mass boson stars for several values of $\gamma$: scalar field at the origin, frequency, mass, charge and compactness. We remark that the numerical computation requires to determine $\omega$ with many more digits of precision than those we are showing.}
\label{t1}
\end{table}

\section{Energy and mass scales}\label{sec:energy}
In the theory \req{theory}, we expressed the scalar field in Planck units for convenience, and as a result, the curvature scale of the sigma model, $\gamma$, is a dimensionless number. However, in order to obtain a more physically correct picture, one should work with the canonically normalized scalar field, $\hat\Phi=\frac{\Phi}{\sqrt{16\pi G}}$, which has dimensions of energy.  As a consequence, its scalar manifold has a dimensionful curvature $k$ corresponding to some characteristic energy scale, $|k|=E^{-2}$.  The parameter $\gamma$ of our models is related to the dimensionful curvature by 
\begin{equation}
\gamma^2=-\frac{k}{16 \pi G}\, .
\end{equation}
Or, in terms of the energy scales, 
\begin{equation}
|\gamma|=\frac{M_{\rm P}}{\sqrt{16\pi} E}\, .
\end{equation}
Now, notice that, in general, the energy scale $E$ of the nonlinear sigma model need not coincide with Planck scale, and it is very natural to assume that $E\ll M_{\rm P}$. Therefore, a large value of $|\gamma|$, like the ones we have studied, is indeed a natural possibility. Writing our formulas \req{mmaxo3} and \req{mmaxsl2} for the maximum mass in terms of the energy scale $E$,  we get
\begin{equation}\label{eq:scalemass}
M_{\rm max}^{\rm{\small{O(3)}}}=0.0798 \frac{M_{\rm P}^3}{\mu E}\, ,\quad 
M_{\rm max}^{\rm{\small{ SL(2,\mathbb{R})}}}=7.8\frac{M_{\rm P}E}{\mu}\, .
\end{equation}
These relations are important as they tell us the range of energy scales for which these models give rise to astrophysical-size boson stars. The $\mathrm{O(3)}$ boson stars have masses comparable to the Chandrasekhar  limit of fermions with mass $m_{f}^2\sim E\mu$. Thus, these solutions can reach stelar masses for a wide range of values of $E$ and $\mu$.  For instance $E\sim\mu\lesssim 1$GeV would yield solar-mass boson stars. 
However, in the $\mathrm{SL}(2,\mathbb{R})$ case with $E\sim\mu$, the maximum mass is Planckian, hence making this model unviable to describe massive compact objects. 
The model could nevertheless still be viable if $E\gg \mu$ and the bosons are ultralight, and especially if $E\sim M_{\rm P}$ (so that $\gamma\sim 1$, as usually happens in string theory and supergravity). In that situation the $\mathrm{SL}(2,\mathbb{R})$ stars would just be comparable to mini boson stars, although somewhat less massive and compact. 

Interestingly, the radius of (maximum mass) boson stars is of the same order of magnitude in both models, and roughly given by $R\sim M_{\rm P}/(E\mu)$. We note that this is much larger than the Compton wavelength of the bosons if $E\ll M_{\rm P}$.  The huge difference in masses means that, within the same radius, the $\mathrm{O(3)}$ model describes almost ultra-compact objects, while the $\mathrm{SL}(2,\mathbb{R})$ stars can be arbitrarily diffuse. 

\section{Discussion}\label{sec:discussion}
We have studied boson star solutions in the $\mathrm{O}(3)$ and $\mathrm{SL}(2,\mathbb{R})$ sigma models. We have considered the case of minimally interacting potentials \req{eq:potentialPHI} in order to focus on the effect of the sigma model curvature on the boson stars. Our results are clear: positive curvature ($\mathrm{O}(3)$) leads to massive and very compact solutions, while negative curvature  ($\mathrm{SL}(2,\mathbb{R})$) produces light and (relatively) diffuse objects. In fact, the masses and compactnesses of these objects differ by many orders of magnitude if the energy scale of the nonlinear sigma model is below Planck scale --- see \req{eq:scalemass} --- but the differences are important even when $E\sim M_{\rm P}$. Therefore, the main conclusion is that the curvature of the sigma model is a crucial factor (as important as the potential) in determining the properties of boson stars. 
While we focused on maximally symmetric sigma models, it would of course be interesting to study how this general conclusion translates to sigma models with non-constant curvature. These, perhaps, could lead to even more drastic effects. 

The case of $\mathrm{O}(3)$ boson stars is particularly interesting, as these objects can acquire very high compactnesses even with a minimal potential. Indeed, our analysis showed that there are presumably stable solutions with a maximum compactness of $R\sim 3.29 GM$. For comparison, this is almost twice the maximum compactness of (stable) massive boson stars \cite{Amaro-Seoane:2010pks} and similar to that of solitonic boson stars \cite{Cardoso:2021ehg}, which are some of the most compact examples known in the literature. 
These $\mathrm{O}(3)$ boson stars are on the verge of becoming ultracompact, that is, developing light rings. Although we did not find solutions with light rings in the naively stable branch, these could perhaps exist for even larger values of the $\mathrm{O}(3)$ model curvature, which become increasingly difficult to explore numerically.\footnote{On the other hand, it is interesting that despite being so compact these objects do not acquire light rings, since these are known to generate a nonlinear instability in horizonless objects \cite{Cunha:2022gde}.} It would be otherwise interesting to explore if other potentials --- including self-interactions --- could increase the compactness of the solutions even more. Likewise, one may wonder if in the case of $\mathrm{SL}(2,\mathbb{R})$ stars, interacting potentials may increase their masses in a substantial way. 

The next natural step in the study of these boson stars would entail the analysis of their linear perturbations and stability. As we mentioned, the maximum mass boson star usually marks the limit of existence of stable solutions, but since these models have a non-standard kinetic term it would be important to investigate if this is still the case. The study of perturbations is furthermore interesting in order to determine observational properties of boson stars, such as their quasinormal modes \cite{Macedo:2013jja} or tidal Love numbers \cite{Mendes:2016vdr,Cardoso:2017cfl}.   On the other hand, studying the image of these boson stars, and determining to what extent they can mimic a black hole, would also prove interesting \cite{Herdeiro:2021lwl, Rosa:2022toh, Rosa:2022tfv, Rosa:2023qcv}. 

Another obvious extension of this work would be studying rotating boson stars, some examples of which were already found in  Ref.~\cite{Herdeiro:2018djx} in the case of $\mathrm{O}(3)$ stars. It would again be interesting to focus on the regime of large curvature of the sigma model in order to investigate if rotating solutions acquire special features in that case.

Finally, one of our motivations to study nonlinear sigma models was their connection to supergravity and string theory. Although the models we have studied can be part of a supergravity action, they do not capture the whole story yet. 
In a supergravity version of these models, there would be a few differences. On the one hand, the unbroken $\mathrm{U}(1)$ symmetry  would typically be gauged, and thus boson stars would be charged \cite{JETZER1989341} hence generating electromagnetic fields.   
On the other hand, the scalar potential, the sigma model metric and the coupling between the scalar and vector fields would be constrained by supersymmetry.  Thus, supergravity would provide a rich (yet restricted) set of theories in which boson star solutions arise naturally. 
The study of these solutions --- which connect some of our most fundamental theories with objects of astrophysical interest\footnote{They would also have a theoretical interest, as boson stars play a role in string theory through the AdS/CFT correspondence \cite{Buchel:2013uba, Buchel:2015rwa, Buchel:2015sma, Milekhin:2023was}.} --- poses a promising avenue of research.    
Our present paper represents the first step in this direction.












\vspace{0.4cm}
\begin{acknowledgments}   
\textbf{\textit{Acknowledgments.}}
PAC would like to thank Alex Buchel for useful comments on the status of boson stars in supergravity. LM would like to thank Gastón Creci for insightful discussion. The work of PAC was supported by a postdoctoral fellowship from the Research Foundation - Flanders (FWO grant 12ZH121N). The project that gave rise to these results received the support of a fellowship from “la Caixa” Foundation (ID 100010434) with code LCF/BQ/PI23/11970032. 
LM is currently a PhD fellow at the Research Foundation - Flanders (FWO grant 1186024N). LM acknowledges funding from ESA Prodex project ’LISA EMRI/IMRAC waveform modelling’ PEA 4000131558, which financed earlier stages of this research.
\end{acknowledgments}

\appendix
 \vspace{1.5cm}

\section{All two-dimensional sigma models with a $U(1)$ isometry}\label{app:sigmamodel}
Without loss of generality, a two-dimensional nonlinear sigma model can always be written as
\begin{equation}\label{eq:invariantLag1}
\mathcal{L}=-H(\tau,\bar\tau)\partial_{\mu}\tau\partial^{\mu}\bar{\tau}-V(\tau,\bar\tau)\, ,
\end{equation}
for a complex scalar $\tau$ and real functions $H(\tau,\bar\tau)$ and $V(\tau,\bar\tau)$. Now we look for theories with a global symmetry, so let us assume that this Lagrangian is invariant under a holomorphic isometry, 
\begin{equation}\label{eq:tranformationtaugeneral}
\tau'=f(\tau;\alpha)\, ,\quad \bar\tau'=\bar{f}(\bar\tau;\alpha)\, ,
\end{equation}
where $f$ is a certain holomorphic function and $\alpha$ is a real parameter labeling the transformation, such that $f(\tau;0)=\tau$. In addition, we are interested in the case in which the symmetry group is $\mathrm{U}(1)$, so 
\begin{equation}
f(\tau;\alpha+2\pi)=f(\tau;\alpha)\, .
\end{equation}
Let us introduce the generators of the symmetry
\begin{equation}
p=\frac{\partial f}{\partial\tau}\bigg|_{\alpha=0}\, ,\quad \bar p=\frac{\partial \bar f}{\partial\bar\tau}\bigg|_{\alpha=0}\, .
\end{equation}
Then, in order for the transformation \req{eq:tranformationtaugeneral} to be a symmetry, it follows that $H$ and $V$ must satisfy
\begin{equation}\label{eq:transfKV}
\begin{aligned}
\frac{\partial H}{\partial\tau}p+\frac{\partial H}{\partial\bar\tau}\bar p+H\left(\frac{\partial p}{\partial \tau}+\frac{\partial \bar p}{\partial \bar \tau}\right)&=0\, ,\\
\frac{\partial V}{\partial\tau}p+\frac{\partial V}{\partial\bar\tau}\bar p&=0\, .
\end{aligned}
\end{equation}

Under these conditions there is a canonical way of writing the Lagrangian. The idea is to introduce a new field $\Phi$ which transforms as 
\begin{equation}\label{eq:Phitransf}
\Phi'=e^{i\alpha}\Phi\, .
\end{equation}
To find the appropriate change of variables $\Phi(\tau)$, we note that under an infinitesimal transformation we have
\begin{equation}
\delta_{\alpha}\Phi=\frac{d\Phi}{d \tau}\delta_{\alpha}\tau=\frac{d\Phi}{d\tau}p\, ,
\end{equation}
and hence demanding $\delta_{\alpha}\Phi=i\Phi$ leads to
\begin{equation}
\frac{d\Phi}{d\tau}p=i\Phi\, .
\end{equation}
This is a differential equation from where we can obtain $\Phi(\tau)$ or the inverse relation $\tau(\Phi)$. It remains to study the form of the Lagrangian when expressed in terms of $\Phi$. We have
\begin{equation}
\mathcal{L}=-K\partial_{\mu}\Phi \partial^{\mu}\bar{\Phi}-V\, ,
\end{equation}
where
\begin{equation}
K=\frac{|p|^2}{|\Phi|^2}H\, ,
\end{equation}
and $V$ is the same as in \req{eq:invariantLag1}, but expressed as a function of $\Phi$. Then, using \req{eq:transfKV}, it follows that
\begin{equation}\label{eq:transfQV}
\begin{aligned}
\Phi\frac{\partial K}{\partial\Phi}-\bar \Phi \frac{\partial K}{\partial\bar\Phi}&=0\, ,\quad 
\Phi\frac{\partial V}{\partial\Phi}-\bar \Phi \frac{\partial V}{\partial\bar\Phi}=0\, ,
\end{aligned}
\end{equation}
hence implying that these quantities only depend on $\Phi$ through its modulus, $K=K\left(|\Phi|^2\right)$, $V=V\left(|\Phi|^2\right)$, as it should be since by construction the transformation \req{eq:Phitransf} should be a symmetry.

Therefore, in terms of the variable $\Phi$, the Lagrangian reads
\begin{equation}
\mathcal{L}=-K\left(|\Phi|^2\right)\partial_{\mu}\Phi\partial^{\mu}\bar{\Phi}-V\left(|\Phi|^2\right)\, .
\end{equation}
While it is obvious that all the theories of this type are invariant under $\mathrm{U}(1)$, it is nontrivial that all nonlinear sigma models with a holomorphic isometry can be expressed in this way.

\section{Numerical methods}\label{app:numerics}

The system of differential equations \req{eq:feq}, \req{eq:geq}, \req{eq:phieq} can be solved by direct integration after the conditions of regularity at the origin and asymptotic flatness are imposed. As discussed in the main text, initial conditions are set at $r_{in}=10^{-3}$ via the expansions \req{eq:BC} and the solution is obtained up to a radius $r_{\rm domain}$, where it diverges for a generic value of $\omega$. However, infinitely supported solutions may exist for specific values of $\omega$, for which the domain of integration rapidly grows.  The goal of the numerical analysis is therefore to identify the peak in $r_{\rm domain}$ corresponding to the resonant frequency of the boson star ground state. With reference to Fig.~\ref{fig:domainr}, this would be leftmost peak. 

First of all, the solutions depend on the choice of the scalar field value at the origin $\phi_0$. For each value of the parameter $\gamma$, there exist a maximum value of $\phi_0$ after which the solutions have bounded $r_{\rm domain}$ for every $\omega$ and asymptotically flat solutions are thus not allowed. As a first step, one needs to find this value $\phi_{0,\rm max}^\gamma$. This is done applying a binary search algorithm on a table of $\phi_0$ values until $\phi_{0,\rm max}^\gamma$ is estimated with sufficient precision. Let us stress that this requires an inefficient search of the first resonant frequency for each value of the central field. This is because in order to assess the existence of resonant solutions, one needs to scan a wide interval of frequencies with a grid  fine enough to correctly identify peaks in $r_{\rm domain}$. Nevertheless, $\phi_{0,\rm max}^\gamma$ can be obtained at a reasonable precision with a low number of binary search iterations.

Once the value of the maximum central field is known for a given value of $\gamma$, we build a table of resonant frequencies for $\phi_0\in[\phi_{0,\rm min}\sim 10^{-2},\phi^\gamma_{0,\rm max}]$. It turns out that for small $\phi_0$, the resonant frequency of the ground state is close to 1. We can therefore fix $\phi_{0, \rm min}$ and start scanning over the frequency from $\omega=1$ (in units of $\mu=1$) with a certain step $\Delta\omega$. We then identify the first peak with a hill climbing method. Once the value $r_{\rm domain}$ decreases --- meaning that we have passed the peak --- we change the sign of the frequency step and refine the grid. This allows to rapidly converge to the left-most peak without having to evaluate the solutions over a large interval of frequencies. Once the first peak is found, the value of the scalar field at the origin is taken to be $\phi_0+\Delta\phi$ and the scan in frequency is performed starting from the previously found value $\omega_{\rm res}(\phi_0)$. As discussed in the main text, for large values of $|\gamma|$ and for $\phi_0\sim\phi^\gamma_{0, \rm max}$, the distance between resonant peaks in the frequency space becomes very small. This makes finding the first resonant peak a difficult problem, as one needs to have a step fine enough to resolve it. This issue can be alleviated by dynamically adapting the frequency step, \textit{i.e.} making it smaller with increasing $|\gamma|$ and $\phi_0$. 

Finally, some convergence checks can be performed on the solutions. In particular, both functions $f(r)$ and $g(r)$ should tend to the Schwarzschild lapse function for $r>R$, \textit{i.e.}
\begin{equation}
	f_0\tilde{f}(r),\,g(r)\xrightarrow{r\rightarrow\infty}\left(1-\frac{2M}{r}\right),
\end{equation} 
with $M$ the ADM mass of the spacetime. Thus, we fit the numerical solutions to the Schwarzschild function and compare the best-fit values of $M$. If the values obtained from $f(r)$ and from $g(r)$ differ by less than $0.01\%$ we consider the result of the integration to be trustworthy. Note that this agreement only happens if the scalar field has decayed enough so that we have effectively reached the asymptotically flat region. 
The value of $f_0$ is also obtained via this fitting procedure. 

\appendix

\bibliographystyle{apsrev4-1} 
\bibliography{StringGravity}

\begin{thebibliography}{54}%
\makeatletter
\providecommand \@ifxundefined [1]{%
 \@ifx{#1\undefined}
}%
\providecommand \@ifnum [1]{%
 \ifnum #1\expandafter \@firstoftwo
 \else \expandafter \@secondoftwo
 \fi
}%
\providecommand \@ifx [1]{%
 \ifx #1\expandafter \@firstoftwo
 \else \expandafter \@secondoftwo
 \fi
}%
\providecommand \natexlab [1]{#1}%
\providecommand \enquote  [1]{``#1''}%
\providecommand \bibnamefont  [1]{#1}%
\providecommand \bibfnamefont [1]{#1}%
\providecommand \citenamefont [1]{#1}%
\providecommand \href@noop [0]{\@secondoftwo}%
\providecommand \href [0]{\begingroup \@sanitize@url \@href}%
\providecommand \@href[1]{\@@startlink{#1}\@@href}%
\providecommand \@@href[1]{\endgroup#1\@@endlink}%
\providecommand \@sanitize@url [0]{\catcode `\\12\catcode `\$12\catcode
  `\&12\catcode `\#12\catcode `\^12\catcode `\_12\catcode `\%12\relax}%
\providecommand \@@startlink[1]{}%
\providecommand \@@endlink[0]{}%
\providecommand \url  [0]{\begingroup\@sanitize@url \@url }%
\providecommand \@url [1]{\endgroup\@href {#1}{\urlprefix }}%
\providecommand \urlprefix  [0]{URL }%
\providecommand \Eprint [0]{\href }%
\providecommand \doibase [0]{http://dx.doi.org/}%
\providecommand \selectlanguage [0]{\@gobble}%
\providecommand \bibinfo  [0]{\@secondoftwo}%
\providecommand \bibfield  [0]{\@secondoftwo}%
\providecommand \translation [1]{[#1]}%
\providecommand \BibitemOpen [0]{}%
\providecommand \bibitemStop [0]{}%
\providecommand \bibitemNoStop [0]{.\EOS\space}%
\providecommand \EOS [0]{\spacefactor3000\relax}%
\providecommand \BibitemShut  [1]{\csname bibitem#1\endcsname}%
\let\auto@bib@innerbib\@empty
\bibitem [{\citenamefont {Wheeler}(1955)}]{PhysRev.97.511}%
  \BibitemOpen
  \bibfield  {author} {\bibinfo {author} {\bibfnamefont {J.~A.}\ \bibnamefont
  {Wheeler}},\ }\href {\doibase 10.1103/PhysRev.97.511} {\bibfield  {journal}
  {\bibinfo  {journal} {Phys. Rev.}\ }\textbf {\bibinfo {volume} {97}},\
  \bibinfo {pages} {511} (\bibinfo {year} {1955})}\BibitemShut {NoStop}%
\bibitem [{\citenamefont {Power}\ and\ \citenamefont
  {Wheeler}(1957)}]{RevModPhys.29.480}%
  \BibitemOpen
  \bibfield  {author} {\bibinfo {author} {\bibfnamefont {E.~A.}\ \bibnamefont
  {Power}}\ and\ \bibinfo {author} {\bibfnamefont {J.~A.}\ \bibnamefont
  {Wheeler}},\ }\href {\doibase 10.1103/RevModPhys.29.480} {\bibfield
  {journal} {\bibinfo  {journal} {Rev. Mod. Phys.}\ }\textbf {\bibinfo {volume}
  {29}},\ \bibinfo {pages} {480} (\bibinfo {year} {1957})}\BibitemShut
  {NoStop}%
\bibitem [{\citenamefont {Abbott}\ \emph {et~al.}(2016)\citenamefont {Abbott}
  \emph {et~al.}}]{LIGOScientific:2016aoc}%
  \BibitemOpen
  \bibfield  {author} {\bibinfo {author} {\bibfnamefont {B.~P.}\ \bibnamefont
  {Abbott}} \emph {et~al.} (\bibinfo {collaboration} {LIGO Scientific,
  Virgo}),\ }\href {\doibase 10.1103/PhysRevLett.116.061102} {\bibfield
  {journal} {\bibinfo  {journal} {Phys. Rev. Lett.}\ }\textbf {\bibinfo
  {volume} {116}},\ \bibinfo {pages} {061102} (\bibinfo {year} {2016})},\
  \Eprint {http://arxiv.org/abs/1602.03837} {arXiv:1602.03837 [gr-qc]}
  \BibitemShut {NoStop}%
\bibitem [{\citenamefont {Cardoso}\ and\ \citenamefont
  {Pani}(2017)}]{Cardoso:2017cqb}%
  \BibitemOpen
  \bibfield  {author} {\bibinfo {author} {\bibfnamefont {V.}~\bibnamefont
  {Cardoso}}\ and\ \bibinfo {author} {\bibfnamefont {P.}~\bibnamefont {Pani}},\
  }\href {\doibase 10.1038/s41550-017-0225-y} {\bibfield  {journal} {\bibinfo
  {journal} {Nature Astron.}\ }\textbf {\bibinfo {volume} {1}},\ \bibinfo
  {pages} {586} (\bibinfo {year} {2017})},\ \Eprint
  {http://arxiv.org/abs/1709.01525} {arXiv:1709.01525 [gr-qc]} \BibitemShut
  {NoStop}%
\bibitem [{\citenamefont {Cardoso}\ and\ \citenamefont
  {Pani}(2019)}]{Cardoso:2019rvt}%
  \BibitemOpen
  \bibfield  {author} {\bibinfo {author} {\bibfnamefont {V.}~\bibnamefont
  {Cardoso}}\ and\ \bibinfo {author} {\bibfnamefont {P.}~\bibnamefont {Pani}},\
  }\href {\doibase 10.1007/s41114-019-0020-4} {\bibfield  {journal} {\bibinfo
  {journal} {Living Rev. Rel.}\ }\textbf {\bibinfo {volume} {22}},\ \bibinfo
  {pages} {4} (\bibinfo {year} {2019})},\ \Eprint
  {http://arxiv.org/abs/1904.05363} {arXiv:1904.05363 [gr-qc]} \BibitemShut
  {NoStop}%
\bibitem [{\citenamefont {Morris}\ \emph {et~al.}(1988)\citenamefont {Morris},
  \citenamefont {Thorne},\ and\ \citenamefont {Yurtsever}}]{Morris:1988tu}%
  \BibitemOpen
  \bibfield  {author} {\bibinfo {author} {\bibfnamefont {M.~S.}\ \bibnamefont
  {Morris}}, \bibinfo {author} {\bibfnamefont {K.~S.}\ \bibnamefont {Thorne}},
  \ and\ \bibinfo {author} {\bibfnamefont {U.}~\bibnamefont {Yurtsever}},\
  }\href {\doibase 10.1103/PhysRevLett.61.1446} {\bibfield  {journal} {\bibinfo
   {journal} {Phys. Rev. Lett.}\ }\textbf {\bibinfo {volume} {61}},\ \bibinfo
  {pages} {1446} (\bibinfo {year} {1988})}\BibitemShut {NoStop}%
\bibitem [{\citenamefont {Mazur}\ and\ \citenamefont
  {Mottola}(2023)}]{Mazur:2001fv}%
  \BibitemOpen
  \bibfield  {author} {\bibinfo {author} {\bibfnamefont {P.~O.}\ \bibnamefont
  {Mazur}}\ and\ \bibinfo {author} {\bibfnamefont {E.}~\bibnamefont
  {Mottola}},\ }\href {\doibase 10.3390/universe9020088} {\bibfield  {journal}
  {\bibinfo  {journal} {Universe}\ }\textbf {\bibinfo {volume} {9}},\ \bibinfo
  {pages} {88} (\bibinfo {year} {2023})},\ \Eprint
  {http://arxiv.org/abs/gr-qc/0109035} {arXiv:gr-qc/0109035} \BibitemShut
  {NoStop}%
\bibitem [{\citenamefont {Mathur}(2005)}]{Mathur:2005zp}%
  \BibitemOpen
  \bibfield  {author} {\bibinfo {author} {\bibfnamefont {S.~D.}\ \bibnamefont
  {Mathur}},\ }\href {\doibase 10.1002/prop.200410203} {\bibfield  {journal}
  {\bibinfo  {journal} {Fortsch. Phys.}\ }\textbf {\bibinfo {volume} {53}},\
  \bibinfo {pages} {793} (\bibinfo {year} {2005})},\ \Eprint
  {http://arxiv.org/abs/hep-th/0502050} {arXiv:hep-th/0502050} \BibitemShut
  {NoStop}%
\bibitem [{\citenamefont {Damour}\ and\ \citenamefont
  {Solodukhin}(2007)}]{Damour:2007ap}%
  \BibitemOpen
  \bibfield  {author} {\bibinfo {author} {\bibfnamefont {T.}~\bibnamefont
  {Damour}}\ and\ \bibinfo {author} {\bibfnamefont {S.~N.}\ \bibnamefont
  {Solodukhin}},\ }\href {\doibase 10.1103/PhysRevD.76.024016} {\bibfield
  {journal} {\bibinfo  {journal} {Phys. Rev. D}\ }\textbf {\bibinfo {volume}
  {76}},\ \bibinfo {pages} {024016} (\bibinfo {year} {2007})},\ \Eprint
  {http://arxiv.org/abs/0704.2667} {arXiv:0704.2667 [gr-qc]} \BibitemShut
  {NoStop}%
\bibitem [{\citenamefont {Holdom}\ and\ \citenamefont
  {Ren}(2017)}]{Holdom:2016nek}%
  \BibitemOpen
  \bibfield  {author} {\bibinfo {author} {\bibfnamefont {B.}~\bibnamefont
  {Holdom}}\ and\ \bibinfo {author} {\bibfnamefont {J.}~\bibnamefont {Ren}},\
  }\href {\doibase 10.1103/PhysRevD.95.084034} {\bibfield  {journal} {\bibinfo
  {journal} {Phys. Rev. D}\ }\textbf {\bibinfo {volume} {95}},\ \bibinfo
  {pages} {084034} (\bibinfo {year} {2017})},\ \Eprint
  {http://arxiv.org/abs/1612.04889} {arXiv:1612.04889 [gr-qc]} \BibitemShut
  {NoStop}%
\bibitem [{\citenamefont {Bueno}\ \emph {et~al.}(2018)\citenamefont {Bueno},
  \citenamefont {Cano}, \citenamefont {Goelen}, \citenamefont {Hertog},\ and\
  \citenamefont {Vercnocke}}]{Bueno:2017hyj}%
  \BibitemOpen
  \bibfield  {author} {\bibinfo {author} {\bibfnamefont {P.}~\bibnamefont
  {Bueno}}, \bibinfo {author} {\bibfnamefont {P.~A.}\ \bibnamefont {Cano}},
  \bibinfo {author} {\bibfnamefont {F.}~\bibnamefont {Goelen}}, \bibinfo
  {author} {\bibfnamefont {T.}~\bibnamefont {Hertog}}, \ and\ \bibinfo {author}
  {\bibfnamefont {B.}~\bibnamefont {Vercnocke}},\ }\href {\doibase
  10.1103/PhysRevD.97.024040} {\bibfield  {journal} {\bibinfo  {journal} {Phys.
  Rev. D}\ }\textbf {\bibinfo {volume} {97}},\ \bibinfo {pages} {024040}
  (\bibinfo {year} {2018})},\ \Eprint {http://arxiv.org/abs/1711.00391}
  {arXiv:1711.00391 [gr-qc]} \BibitemShut {NoStop}%
\bibitem [{\citenamefont {Mayerson}\ and\ \citenamefont
  {Vercnocke}(2023)}]{Mayerson:2023wck}%
  \BibitemOpen
  \bibfield  {author} {\bibinfo {author} {\bibfnamefont {D.~R.}\ \bibnamefont
  {Mayerson}}\ and\ \bibinfo {author} {\bibfnamefont {B.}~\bibnamefont
  {Vercnocke}},\ }\href@noop {} {\  (\bibinfo {year} {2023})},\ \Eprint
  {http://arxiv.org/abs/2306.01565} {arXiv:2306.01565 [hep-th]} \BibitemShut
  {NoStop}%
\bibitem [{\citenamefont {Jetzer}(1992)}]{JETZER1992163}%
  \BibitemOpen
  \bibfield  {author} {\bibinfo {author} {\bibfnamefont {P.}~\bibnamefont
  {Jetzer}},\ }\href {\doibase https://doi.org/10.1016/0370-1573(92)90123-H}
  {\bibfield  {journal} {\bibinfo  {journal} {Physics Reports}\ }\textbf
  {\bibinfo {volume} {220}},\ \bibinfo {pages} {163} (\bibinfo {year}
  {1992})}\BibitemShut {NoStop}%
\bibitem [{\citenamefont {Liebling}\ and\ \citenamefont
  {Palenzuela}(2023)}]{Liebling:2012fv}%
  \BibitemOpen
  \bibfield  {author} {\bibinfo {author} {\bibfnamefont {S.~L.}\ \bibnamefont
  {Liebling}}\ and\ \bibinfo {author} {\bibfnamefont {C.}~\bibnamefont
  {Palenzuela}},\ }\href {\doibase 10.1007/s41114-023-00043-4} {\bibfield
  {journal} {\bibinfo  {journal} {Living Rev. Rel.}\ }\textbf {\bibinfo
  {volume} {26}},\ \bibinfo {pages} {1} (\bibinfo {year} {2023})},\ \Eprint
  {http://arxiv.org/abs/1202.5809} {arXiv:1202.5809 [gr-qc]} \BibitemShut
  {NoStop}%
\bibitem [{\citenamefont {Visinelli}(2021)}]{Visinelli:2021uve}%
  \BibitemOpen
  \bibfield  {author} {\bibinfo {author} {\bibfnamefont {L.}~\bibnamefont
  {Visinelli}},\ }\href {\doibase 10.1142/S0218271821300068} {\bibfield
  {journal} {\bibinfo  {journal} {Int. J. Mod. Phys. D}\ }\textbf {\bibinfo
  {volume} {30}},\ \bibinfo {pages} {2130006} (\bibinfo {year} {2021})},\
  \Eprint {http://arxiv.org/abs/2109.05481} {arXiv:2109.05481 [gr-qc]}
  \BibitemShut {NoStop}%
\bibitem [{\citenamefont {Lee}\ and\ \citenamefont {Pang}(1989)}]{Lee:1988av}%
  \BibitemOpen
  \bibfield  {author} {\bibinfo {author} {\bibfnamefont {T.~D.}\ \bibnamefont
  {Lee}}\ and\ \bibinfo {author} {\bibfnamefont {Y.}~\bibnamefont {Pang}},\
  }\href {\doibase 10.1016/0550-3213(89)90365-9} {\bibfield  {journal}
  {\bibinfo  {journal} {Nucl. Phys. B}\ }\textbf {\bibinfo {volume} {315}},\
  \bibinfo {pages} {477} (\bibinfo {year} {1989})}\BibitemShut {NoStop}%
\bibitem [{\citenamefont {Gleiser}(1988)}]{Gleiser:1988rq}%
  \BibitemOpen
  \bibfield  {author} {\bibinfo {author} {\bibfnamefont {M.}~\bibnamefont
  {Gleiser}},\ }\href {\doibase 10.1103/PhysRevD.38.2376} {\bibfield  {journal}
  {\bibinfo  {journal} {Phys. Rev. D}\ }\textbf {\bibinfo {volume} {38}},\
  \bibinfo {pages} {2376} (\bibinfo {year} {1988})},\ \bibinfo {note}
  {[Erratum: Phys.Rev.D 39, 1257 (1989)]}\BibitemShut {NoStop}%
\bibitem [{\citenamefont {Schunck}\ and\ \citenamefont
  {Mielke}(2003)}]{Schunck:2003kk}%
  \BibitemOpen
  \bibfield  {author} {\bibinfo {author} {\bibfnamefont {F.~E.}\ \bibnamefont
  {Schunck}}\ and\ \bibinfo {author} {\bibfnamefont {E.~W.}\ \bibnamefont
  {Mielke}},\ }\href {\doibase 10.1088/0264-9381/20/20/201} {\bibfield
  {journal} {\bibinfo  {journal} {Class. Quant. Grav.}\ }\textbf {\bibinfo
  {volume} {20}},\ \bibinfo {pages} {R301} (\bibinfo {year} {2003})},\ \Eprint
  {http://arxiv.org/abs/0801.0307} {arXiv:0801.0307 [astro-ph]} \BibitemShut
  {NoStop}%
\bibitem [{\citenamefont {Macedo}\ \emph
  {et~al.}(2013{\natexlab{a}})\citenamefont {Macedo}, \citenamefont {Pani},
  \citenamefont {Cardoso},\ and\ \citenamefont {Crispino}}]{Macedo:2013qea}%
  \BibitemOpen
  \bibfield  {author} {\bibinfo {author} {\bibfnamefont {C.~F.~B.}\
  \bibnamefont {Macedo}}, \bibinfo {author} {\bibfnamefont {P.}~\bibnamefont
  {Pani}}, \bibinfo {author} {\bibfnamefont {V.}~\bibnamefont {Cardoso}}, \
  and\ \bibinfo {author} {\bibfnamefont {L.~C.~B.}\ \bibnamefont {Crispino}},\
  }\href {\doibase 10.1088/0004-637X/774/1/48} {\bibfield  {journal} {\bibinfo
  {journal} {Astrophys. J.}\ }\textbf {\bibinfo {volume} {774}},\ \bibinfo
  {pages} {48} (\bibinfo {year} {2013}{\natexlab{a}})},\ \Eprint
  {http://arxiv.org/abs/1302.2646} {arXiv:1302.2646 [gr-qc]} \BibitemShut
  {NoStop}%
\bibitem [{\citenamefont {Kaup}(1968)}]{PhysRev.172.1331}%
  \BibitemOpen
  \bibfield  {author} {\bibinfo {author} {\bibfnamefont {D.~J.}\ \bibnamefont
  {Kaup}},\ }\href {\doibase 10.1103/PhysRev.172.1331} {\bibfield  {journal}
  {\bibinfo  {journal} {Phys. Rev.}\ }\textbf {\bibinfo {volume} {172}},\
  \bibinfo {pages} {1331} (\bibinfo {year} {1968})}\BibitemShut {NoStop}%
\bibitem [{\citenamefont {RUFFINI}\ and\ \citenamefont
  {BONAZZOLA}(1969)}]{PhysRev.187.1767}%
  \BibitemOpen
  \bibfield  {author} {\bibinfo {author} {\bibfnamefont {R.}~\bibnamefont
  {RUFFINI}}\ and\ \bibinfo {author} {\bibfnamefont {S.}~\bibnamefont
  {BONAZZOLA}},\ }\href {\doibase 10.1103/PhysRev.187.1767} {\bibfield
  {journal} {\bibinfo  {journal} {Phys. Rev.}\ }\textbf {\bibinfo {volume}
  {187}},\ \bibinfo {pages} {1767} (\bibinfo {year} {1969})}\BibitemShut
  {NoStop}%
\bibitem [{\citenamefont {Colpi}\ \emph {et~al.}(1986)\citenamefont {Colpi},
  \citenamefont {Shapiro},\ and\ \citenamefont
  {Wasserman}}]{PhysRevLett.57.2485}%
  \BibitemOpen
  \bibfield  {author} {\bibinfo {author} {\bibfnamefont {M.}~\bibnamefont
  {Colpi}}, \bibinfo {author} {\bibfnamefont {S.~L.}\ \bibnamefont {Shapiro}},
  \ and\ \bibinfo {author} {\bibfnamefont {I.}~\bibnamefont {Wasserman}},\
  }\href {\doibase 10.1103/PhysRevLett.57.2485} {\bibfield  {journal} {\bibinfo
   {journal} {Phys. Rev. Lett.}\ }\textbf {\bibinfo {volume} {57}},\ \bibinfo
  {pages} {2485} (\bibinfo {year} {1986})}\BibitemShut {NoStop}%
\bibitem [{\citenamefont {Khlopov}\ \emph {et~al.}(1985)\citenamefont
  {Khlopov}, \citenamefont {Malomed},\ and\ \citenamefont
  {Zeldovich}}]{Khlopov:1985jw}%
  \BibitemOpen
  \bibfield  {author} {\bibinfo {author} {\bibfnamefont {M.}~\bibnamefont
  {Khlopov}}, \bibinfo {author} {\bibfnamefont {B.~A.}\ \bibnamefont
  {Malomed}}, \ and\ \bibinfo {author} {\bibfnamefont {I.~B.}\ \bibnamefont
  {Zeldovich}},\ }\href@noop {} {\bibfield  {journal} {\bibinfo  {journal}
  {Mon. Not. Roy. Astron. Soc.}\ }\textbf {\bibinfo {volume} {215}},\ \bibinfo
  {pages} {575} (\bibinfo {year} {1985})}\BibitemShut {NoStop}%
\bibitem [{\citenamefont {Lee}(1987)}]{PhysRevD.35.3637}%
  \BibitemOpen
  \bibfield  {author} {\bibinfo {author} {\bibfnamefont {T.~D.}\ \bibnamefont
  {Lee}},\ }\href {\doibase 10.1103/PhysRevD.35.3637} {\bibfield  {journal}
  {\bibinfo  {journal} {Phys. Rev. D}\ }\textbf {\bibinfo {volume} {35}},\
  \bibinfo {pages} {3637} (\bibinfo {year} {1987})}\BibitemShut {NoStop}%
\bibitem [{\citenamefont {Friedberg}\ \emph {et~al.}(1987)\citenamefont
  {Friedberg}, \citenamefont {Lee},\ and\ \citenamefont
  {Pang}}]{PhysRevD.35.3658}%
  \BibitemOpen
  \bibfield  {author} {\bibinfo {author} {\bibfnamefont {R.}~\bibnamefont
  {Friedberg}}, \bibinfo {author} {\bibfnamefont {T.~D.}\ \bibnamefont {Lee}},
  \ and\ \bibinfo {author} {\bibfnamefont {Y.}~\bibnamefont {Pang}},\ }\href
  {\doibase 10.1103/PhysRevD.35.3658} {\bibfield  {journal} {\bibinfo
  {journal} {Phys. Rev. D}\ }\textbf {\bibinfo {volume} {35}},\ \bibinfo
  {pages} {3658} (\bibinfo {year} {1987})}\BibitemShut {NoStop}%
\bibitem [{\citenamefont {Lee}\ and\ \citenamefont {Pang}(1992)}]{LEE1992251}%
  \BibitemOpen
  \bibfield  {author} {\bibinfo {author} {\bibfnamefont {T.}~\bibnamefont
  {Lee}}\ and\ \bibinfo {author} {\bibfnamefont {Y.}~\bibnamefont {Pang}},\
  }\href {\doibase https://doi.org/10.1016/0370-1573(92)90064-7} {\bibfield
  {journal} {\bibinfo  {journal} {Physics Reports}\ }\textbf {\bibinfo {volume}
  {221}},\ \bibinfo {pages} {251} (\bibinfo {year} {1992})}\BibitemShut
  {NoStop}%
\bibitem [{\citenamefont {Guerra}\ \emph {et~al.}(2019)\citenamefont {Guerra},
  \citenamefont {Macedo},\ and\ \citenamefont {Pani}}]{Guerra:2019srj}%
  \BibitemOpen
  \bibfield  {author} {\bibinfo {author} {\bibfnamefont {D.}~\bibnamefont
  {Guerra}}, \bibinfo {author} {\bibfnamefont {C.~F.~B.}\ \bibnamefont
  {Macedo}}, \ and\ \bibinfo {author} {\bibfnamefont {P.}~\bibnamefont
  {Pani}},\ }\href {\doibase 10.1088/1475-7516/2019/09/061} {\bibfield
  {journal} {\bibinfo  {journal} {JCAP}\ }\textbf {\bibinfo {volume} {09}},\
  \bibinfo {pages} {061} (\bibinfo {year} {2019})},\ \bibinfo {note} {[Erratum:
  JCAP 06, E01 (2020)]},\ \Eprint {http://arxiv.org/abs/1909.05515}
  {arXiv:1909.05515 [gr-qc]} \BibitemShut {NoStop}%
\bibitem [{\citenamefont {Delgado}\ \emph {et~al.}(2020)\citenamefont
  {Delgado}, \citenamefont {Herdeiro},\ and\ \citenamefont
  {Radu}}]{Delgado:2020udb}%
  \BibitemOpen
  \bibfield  {author} {\bibinfo {author} {\bibfnamefont {J.~F.~M.}\
  \bibnamefont {Delgado}}, \bibinfo {author} {\bibfnamefont {C.~A.~R.}\
  \bibnamefont {Herdeiro}}, \ and\ \bibinfo {author} {\bibfnamefont
  {E.}~\bibnamefont {Radu}},\ }\href {\doibase 10.1088/1475-7516/2020/06/037}
  {\bibfield  {journal} {\bibinfo  {journal} {JCAP}\ }\textbf {\bibinfo
  {volume} {06}},\ \bibinfo {pages} {037} (\bibinfo {year} {2020})},\ \Eprint
  {http://arxiv.org/abs/2005.05982} {arXiv:2005.05982 [gr-qc]} \BibitemShut
  {NoStop}%
\bibitem [{\citenamefont {Alcubierre}\ \emph {et~al.}(2018)\citenamefont
  {Alcubierre}, \citenamefont {Barranco}, \citenamefont {Bernal}, \citenamefont
  {Degollado}, \citenamefont {Diez-Tejedor}, \citenamefont {Megevand},
  \citenamefont {Nunez},\ and\ \citenamefont {Sarbach}}]{Alcubierre:2018ahf}%
  \BibitemOpen
  \bibfield  {author} {\bibinfo {author} {\bibfnamefont {M.}~\bibnamefont
  {Alcubierre}}, \bibinfo {author} {\bibfnamefont {J.}~\bibnamefont
  {Barranco}}, \bibinfo {author} {\bibfnamefont {A.}~\bibnamefont {Bernal}},
  \bibinfo {author} {\bibfnamefont {J.~C.}\ \bibnamefont {Degollado}}, \bibinfo
  {author} {\bibfnamefont {A.}~\bibnamefont {Diez-Tejedor}}, \bibinfo {author}
  {\bibfnamefont {M.}~\bibnamefont {Megevand}}, \bibinfo {author}
  {\bibfnamefont {D.}~\bibnamefont {Nunez}}, \ and\ \bibinfo {author}
  {\bibfnamefont {O.}~\bibnamefont {Sarbach}},\ }\href {\doibase
  10.1088/1361-6382/aadcb6} {\bibfield  {journal} {\bibinfo  {journal} {Class.
  Quant. Grav.}\ }\textbf {\bibinfo {volume} {35}},\ \bibinfo {pages} {19LT01}
  (\bibinfo {year} {2018})},\ \Eprint {http://arxiv.org/abs/1805.11488}
  {arXiv:1805.11488 [gr-qc]} \BibitemShut {NoStop}%
\bibitem [{\citenamefont {Alcubierre}\ \emph {et~al.}(2019)\citenamefont
  {Alcubierre}, \citenamefont {Barranco}, \citenamefont {Bernal}, \citenamefont
  {Degollado}, \citenamefont {Diez-Tejedor}, \citenamefont {Megevand},
  \citenamefont {N\'u\~nez},\ and\ \citenamefont
  {Sarbach}}]{Alcubierre:2019qnh}%
  \BibitemOpen
  \bibfield  {author} {\bibinfo {author} {\bibfnamefont {M.}~\bibnamefont
  {Alcubierre}}, \bibinfo {author} {\bibfnamefont {J.}~\bibnamefont
  {Barranco}}, \bibinfo {author} {\bibfnamefont {A.}~\bibnamefont {Bernal}},
  \bibinfo {author} {\bibfnamefont {J.~C.}\ \bibnamefont {Degollado}}, \bibinfo
  {author} {\bibfnamefont {A.}~\bibnamefont {Diez-Tejedor}}, \bibinfo {author}
  {\bibfnamefont {M.}~\bibnamefont {Megevand}}, \bibinfo {author}
  {\bibfnamefont {D.}~\bibnamefont {N\'u\~nez}}, \ and\ \bibinfo {author}
  {\bibfnamefont {O.}~\bibnamefont {Sarbach}},\ }\href {\doibase
  10.1088/1361-6382/ab4726} {\bibfield  {journal} {\bibinfo  {journal} {Class.
  Quant. Grav.}\ }\textbf {\bibinfo {volume} {36}},\ \bibinfo {pages} {215013}
  (\bibinfo {year} {2019})},\ \Eprint {http://arxiv.org/abs/1906.08959}
  {arXiv:1906.08959 [gr-qc]} \BibitemShut {NoStop}%
\bibitem [{\citenamefont {Jaramillo}\ \emph {et~al.}(2020)\citenamefont
  {Jaramillo}, \citenamefont {Sanchis-Gual}, \citenamefont {Barranco},
  \citenamefont {Bernal}, \citenamefont {Degollado}, \citenamefont {Herdeiro},\
  and\ \citenamefont {N\'u\~nez}}]{Jaramillo:2020rsv}%
  \BibitemOpen
  \bibfield  {author} {\bibinfo {author} {\bibfnamefont {V.}~\bibnamefont
  {Jaramillo}}, \bibinfo {author} {\bibfnamefont {N.}~\bibnamefont
  {Sanchis-Gual}}, \bibinfo {author} {\bibfnamefont {J.}~\bibnamefont
  {Barranco}}, \bibinfo {author} {\bibfnamefont {A.}~\bibnamefont {Bernal}},
  \bibinfo {author} {\bibfnamefont {J.~C.}\ \bibnamefont {Degollado}}, \bibinfo
  {author} {\bibfnamefont {C.}~\bibnamefont {Herdeiro}}, \ and\ \bibinfo
  {author} {\bibfnamefont {D.}~\bibnamefont {N\'u\~nez}},\ }\href {\doibase
  10.1103/PhysRevD.101.124020} {\bibfield  {journal} {\bibinfo  {journal}
  {Phys. Rev. D}\ }\textbf {\bibinfo {volume} {101}},\ \bibinfo {pages}
  {124020} (\bibinfo {year} {2020})},\ \Eprint
  {http://arxiv.org/abs/2004.08459} {arXiv:2004.08459 [gr-qc]} \BibitemShut
  {NoStop}%
\bibitem [{\citenamefont {Alcubierre}\ \emph {et~al.}(2022)\citenamefont
  {Alcubierre}, \citenamefont {Barranco}, \citenamefont {Bernal}, \citenamefont
  {Degollado}, \citenamefont {Diez-Tejedor}, \citenamefont {Jaramillo},
  \citenamefont {Megevand}, \citenamefont {N\'u\~nez},\ and\ \citenamefont
  {Sarbach}}]{Alcubierre:2021psa}%
  \BibitemOpen
  \bibfield  {author} {\bibinfo {author} {\bibfnamefont {M.}~\bibnamefont
  {Alcubierre}}, \bibinfo {author} {\bibfnamefont {J.}~\bibnamefont
  {Barranco}}, \bibinfo {author} {\bibfnamefont {A.}~\bibnamefont {Bernal}},
  \bibinfo {author} {\bibfnamefont {J.~C.}\ \bibnamefont {Degollado}}, \bibinfo
  {author} {\bibfnamefont {A.}~\bibnamefont {Diez-Tejedor}}, \bibinfo {author}
  {\bibfnamefont {V.}~\bibnamefont {Jaramillo}}, \bibinfo {author}
  {\bibfnamefont {M.}~\bibnamefont {Megevand}}, \bibinfo {author}
  {\bibfnamefont {D.}~\bibnamefont {N\'u\~nez}}, \ and\ \bibinfo {author}
  {\bibfnamefont {O.}~\bibnamefont {Sarbach}},\ }\href {\doibase
  10.1088/1361-6382/ac5fc2} {\bibfield  {journal} {\bibinfo  {journal} {Class.
  Quant. Grav.}\ }\textbf {\bibinfo {volume} {39}},\ \bibinfo {pages} {094001}
  (\bibinfo {year} {2022})},\ \Eprint {http://arxiv.org/abs/2112.04529}
  {arXiv:2112.04529 [gr-qc]} \BibitemShut {NoStop}%
\bibitem [{\citenamefont {Mielke}(2016)}]{Mielke:2016war}%
  \BibitemOpen
  \bibfield  {author} {\bibinfo {author} {\bibfnamefont {E.~W.}\ \bibnamefont
  {Mielke}},\ }\href {\doibase 10.1007/978-3-319-31299-6_6} {\bibfield
  {journal} {\bibinfo  {journal} {Fundam. Theor. Phys.}\ }\textbf {\bibinfo
  {volume} {183}},\ \bibinfo {pages} {115} (\bibinfo {year}
  {2016})}\BibitemShut {NoStop}%
\bibitem [{\citenamefont {Ortin}(2015)}]{Ortin:2015hya}%
  \BibitemOpen
  \bibfield  {author} {\bibinfo {author} {\bibfnamefont {T.}~\bibnamefont
  {Ortin}},\ }\href {\doibase 10.1017/CBO9781139019750} {\emph {\bibinfo
  {title} {{Gravity and Strings}}}},\ \bibinfo {edition} {2nd}\ ed.,\ Cambridge
  Monographs on Mathematical Physics\ (\bibinfo  {publisher} {Cambridge
  University Press},\ \bibinfo {year} {2015})\BibitemShut {NoStop}%
\bibitem [{\citenamefont {Ho}\ \emph {et~al.}(1999)\citenamefont {Ho},
  \citenamefont {Kim},\ and\ \citenamefont {Lee}}]{Ho:1999hs}%
  \BibitemOpen
  \bibfield  {author} {\bibinfo {author} {\bibfnamefont {J.}~\bibnamefont
  {Ho}}, \bibinfo {author} {\bibfnamefont {S.-j.}\ \bibnamefont {Kim}}, \ and\
  \bibinfo {author} {\bibfnamefont {B.-H.}\ \bibnamefont {Lee}},\ }\href@noop
  {} {\  (\bibinfo {year} {1999})},\ \Eprint
  {http://arxiv.org/abs/gr-qc/9902040} {arXiv:gr-qc/9902040} \BibitemShut
  {NoStop}%
\bibitem [{\citenamefont {Verbin}(2007)}]{Verbin:2007fa}%
  \BibitemOpen
  \bibfield  {author} {\bibinfo {author} {\bibfnamefont {Y.}~\bibnamefont
  {Verbin}},\ }\href {\doibase 10.1103/PhysRevD.76.085018} {\bibfield
  {journal} {\bibinfo  {journal} {Phys. Rev. D}\ }\textbf {\bibinfo {volume}
  {76}},\ \bibinfo {pages} {085018} (\bibinfo {year} {2007})},\ \Eprint
  {http://arxiv.org/abs/0708.3283} {arXiv:0708.3283 [gr-qc]} \BibitemShut
  {NoStop}%
\bibitem [{\citenamefont {Herdeiro}\ \emph {et~al.}(2019)\citenamefont
  {Herdeiro}, \citenamefont {Perapechka}, \citenamefont {Radu},\ and\
  \citenamefont {Shnir}}]{Herdeiro:2018djx}%
  \BibitemOpen
  \bibfield  {author} {\bibinfo {author} {\bibfnamefont {C.}~\bibnamefont
  {Herdeiro}}, \bibinfo {author} {\bibfnamefont {I.}~\bibnamefont
  {Perapechka}}, \bibinfo {author} {\bibfnamefont {E.}~\bibnamefont {Radu}}, \
  and\ \bibinfo {author} {\bibfnamefont {Y.}~\bibnamefont {Shnir}},\ }\href
  {\doibase 10.1007/JHEP02(2019)111} {\bibfield  {journal} {\bibinfo  {journal}
  {JHEP}\ }\textbf {\bibinfo {volume} {02}},\ \bibinfo {pages} {111} (\bibinfo
  {year} {2019})},\ \Eprint {http://arxiv.org/abs/1811.11799} {arXiv:1811.11799
  [gr-qc]} \BibitemShut {NoStop}%
\bibitem [{\citenamefont {Mielke}\ and\ \citenamefont
  {Schunck}(2001)}]{Mielke:2001td}%
  \BibitemOpen
  \bibfield  {author} {\bibinfo {author} {\bibfnamefont {E.~W.}\ \bibnamefont
  {Mielke}}\ and\ \bibinfo {author} {\bibfnamefont {F.~E.}\ \bibnamefont
  {Schunck}},\ }\href {\doibase 10.1023/A:1010251707026} {\bibfield  {journal}
  {\bibinfo  {journal} {Gen. Rel. Grav.}\ }\textbf {\bibinfo {volume} {33}},\
  \bibinfo {pages} {805} (\bibinfo {year} {2001})}\BibitemShut {NoStop}%
\bibitem [{\citenamefont {Cunha}\ \emph {et~al.}(2017)\citenamefont {Cunha},
  \citenamefont {Font}, \citenamefont {Herdeiro}, \citenamefont {Radu},
  \citenamefont {Sanchis-Gual},\ and\ \citenamefont
  {Zilh\~ao}}]{Cunha:2017wao}%
  \BibitemOpen
  \bibfield  {author} {\bibinfo {author} {\bibfnamefont {P.~V.~P.}\
  \bibnamefont {Cunha}}, \bibinfo {author} {\bibfnamefont {J.~A.}\ \bibnamefont
  {Font}}, \bibinfo {author} {\bibfnamefont {C.}~\bibnamefont {Herdeiro}},
  \bibinfo {author} {\bibfnamefont {E.}~\bibnamefont {Radu}}, \bibinfo {author}
  {\bibfnamefont {N.}~\bibnamefont {Sanchis-Gual}}, \ and\ \bibinfo {author}
  {\bibfnamefont {M.}~\bibnamefont {Zilh\~ao}},\ }\href {\doibase
  10.1103/PhysRevD.96.104040} {\bibfield  {journal} {\bibinfo  {journal} {Phys.
  Rev. D}\ }\textbf {\bibinfo {volume} {96}},\ \bibinfo {pages} {104040}
  (\bibinfo {year} {2017})},\ \Eprint {http://arxiv.org/abs/1709.06118}
  {arXiv:1709.06118 [gr-qc]} \BibitemShut {NoStop}%
\bibitem [{\citenamefont {Amaro-Seoane}\ \emph {et~al.}(2010)\citenamefont
  {Amaro-Seoane}, \citenamefont {Barranco}, \citenamefont {Bernal},\ and\
  \citenamefont {Rezzolla}}]{Amaro-Seoane:2010pks}%
  \BibitemOpen
  \bibfield  {author} {\bibinfo {author} {\bibfnamefont {P.}~\bibnamefont
  {Amaro-Seoane}}, \bibinfo {author} {\bibfnamefont {J.}~\bibnamefont
  {Barranco}}, \bibinfo {author} {\bibfnamefont {A.}~\bibnamefont {Bernal}}, \
  and\ \bibinfo {author} {\bibfnamefont {L.}~\bibnamefont {Rezzolla}},\ }\href
  {\doibase 10.1088/1475-7516/2010/11/002} {\bibfield  {journal} {\bibinfo
  {journal} {JCAP}\ }\textbf {\bibinfo {volume} {11}},\ \bibinfo {pages} {002}
  (\bibinfo {year} {2010})},\ \Eprint {http://arxiv.org/abs/1009.0019}
  {arXiv:1009.0019 [astro-ph.CO]} \BibitemShut {NoStop}%
\bibitem [{\citenamefont {Cardoso}\ \emph {et~al.}(2022)\citenamefont
  {Cardoso}, \citenamefont {Macedo}, \citenamefont {Maeda},\ and\ \citenamefont
  {Okawa}}]{Cardoso:2021ehg}%
  \BibitemOpen
  \bibfield  {author} {\bibinfo {author} {\bibfnamefont {V.}~\bibnamefont
  {Cardoso}}, \bibinfo {author} {\bibfnamefont {C.~F.~B.}\ \bibnamefont
  {Macedo}}, \bibinfo {author} {\bibfnamefont {K.-i.}\ \bibnamefont {Maeda}}, \
  and\ \bibinfo {author} {\bibfnamefont {H.}~\bibnamefont {Okawa}},\ }\href
  {\doibase 10.1088/1361-6382/ac41e7} {\bibfield  {journal} {\bibinfo
  {journal} {Class. Quant. Grav.}\ }\textbf {\bibinfo {volume} {39}},\ \bibinfo
  {pages} {034001} (\bibinfo {year} {2022})},\ \Eprint
  {http://arxiv.org/abs/2112.05750} {arXiv:2112.05750 [gr-qc]} \BibitemShut
  {NoStop}%
\bibitem [{\citenamefont {Cunha}\ \emph {et~al.}(2023)\citenamefont {Cunha},
  \citenamefont {Herdeiro}, \citenamefont {Radu},\ and\ \citenamefont
  {Sanchis-Gual}}]{Cunha:2022gde}%
  \BibitemOpen
  \bibfield  {author} {\bibinfo {author} {\bibfnamefont {P.~V.~P.}\
  \bibnamefont {Cunha}}, \bibinfo {author} {\bibfnamefont {C.}~\bibnamefont
  {Herdeiro}}, \bibinfo {author} {\bibfnamefont {E.}~\bibnamefont {Radu}}, \
  and\ \bibinfo {author} {\bibfnamefont {N.}~\bibnamefont {Sanchis-Gual}},\
  }\href {\doibase 10.1103/PhysRevLett.130.061401} {\bibfield  {journal}
  {\bibinfo  {journal} {Phys. Rev. Lett.}\ }\textbf {\bibinfo {volume} {130}},\
  \bibinfo {pages} {061401} (\bibinfo {year} {2023})},\ \Eprint
  {http://arxiv.org/abs/2207.13713} {arXiv:2207.13713 [gr-qc]} \BibitemShut
  {NoStop}%
\bibitem [{\citenamefont {Macedo}\ \emph
  {et~al.}(2013{\natexlab{b}})\citenamefont {Macedo}, \citenamefont {Pani},
  \citenamefont {Cardoso},\ and\ \citenamefont {Crispino}}]{Macedo:2013jja}%
  \BibitemOpen
  \bibfield  {author} {\bibinfo {author} {\bibfnamefont {C.~F.~B.}\
  \bibnamefont {Macedo}}, \bibinfo {author} {\bibfnamefont {P.}~\bibnamefont
  {Pani}}, \bibinfo {author} {\bibfnamefont {V.}~\bibnamefont {Cardoso}}, \
  and\ \bibinfo {author} {\bibfnamefont {L.~C.~B.}\ \bibnamefont {Crispino}},\
  }\href {\doibase 10.1103/PhysRevD.88.064046} {\bibfield  {journal} {\bibinfo
  {journal} {Phys. Rev. D}\ }\textbf {\bibinfo {volume} {88}},\ \bibinfo
  {pages} {064046} (\bibinfo {year} {2013}{\natexlab{b}})},\ \Eprint
  {http://arxiv.org/abs/1307.4812} {arXiv:1307.4812 [gr-qc]} \BibitemShut
  {NoStop}%
\bibitem [{\citenamefont {Mendes}\ and\ \citenamefont
  {Yang}(2017)}]{Mendes:2016vdr}%
  \BibitemOpen
  \bibfield  {author} {\bibinfo {author} {\bibfnamefont {R.~F.~P.}\
  \bibnamefont {Mendes}}\ and\ \bibinfo {author} {\bibfnamefont
  {H.}~\bibnamefont {Yang}},\ }\href {\doibase 10.1088/1361-6382/aa842d}
  {\bibfield  {journal} {\bibinfo  {journal} {Class. Quant. Grav.}\ }\textbf
  {\bibinfo {volume} {34}},\ \bibinfo {pages} {185001} (\bibinfo {year}
  {2017})},\ \Eprint {http://arxiv.org/abs/1606.03035} {arXiv:1606.03035
  [astro-ph.CO]} \BibitemShut {NoStop}%
\bibitem [{\citenamefont {Cardoso}\ \emph {et~al.}(2017)\citenamefont
  {Cardoso}, \citenamefont {Franzin}, \citenamefont {Maselli}, \citenamefont
  {Pani},\ and\ \citenamefont {Raposo}}]{Cardoso:2017cfl}%
  \BibitemOpen
  \bibfield  {author} {\bibinfo {author} {\bibfnamefont {V.}~\bibnamefont
  {Cardoso}}, \bibinfo {author} {\bibfnamefont {E.}~\bibnamefont {Franzin}},
  \bibinfo {author} {\bibfnamefont {A.}~\bibnamefont {Maselli}}, \bibinfo
  {author} {\bibfnamefont {P.}~\bibnamefont {Pani}}, \ and\ \bibinfo {author}
  {\bibfnamefont {G.}~\bibnamefont {Raposo}},\ }\href {\doibase
  10.1103/PhysRevD.95.084014} {\bibfield  {journal} {\bibinfo  {journal} {Phys.
  Rev. D}\ }\textbf {\bibinfo {volume} {95}},\ \bibinfo {pages} {084014}
  (\bibinfo {year} {2017})},\ \bibinfo {note} {[Addendum: Phys.Rev.D 95, 089901
  (2017)]},\ \Eprint {http://arxiv.org/abs/1701.01116} {arXiv:1701.01116
  [gr-qc]} \BibitemShut {NoStop}%
\bibitem [{\citenamefont {Herdeiro}\ \emph {et~al.}(2021)\citenamefont
  {Herdeiro}, \citenamefont {Pombo}, \citenamefont {Radu}, \citenamefont
  {Cunha},\ and\ \citenamefont {Sanchis-Gual}}]{Herdeiro:2021lwl}%
  \BibitemOpen
  \bibfield  {author} {\bibinfo {author} {\bibfnamefont {C.~A.~R.}\
  \bibnamefont {Herdeiro}}, \bibinfo {author} {\bibfnamefont {A.~M.}\
  \bibnamefont {Pombo}}, \bibinfo {author} {\bibfnamefont {E.}~\bibnamefont
  {Radu}}, \bibinfo {author} {\bibfnamefont {P.~V.~P.}\ \bibnamefont {Cunha}},
  \ and\ \bibinfo {author} {\bibfnamefont {N.}~\bibnamefont {Sanchis-Gual}},\
  }\href {\doibase 10.1088/1475-7516/2021/04/051} {\bibfield  {journal}
  {\bibinfo  {journal} {JCAP}\ }\textbf {\bibinfo {volume} {04}},\ \bibinfo
  {pages} {051} (\bibinfo {year} {2021})},\ \Eprint
  {http://arxiv.org/abs/2102.01703} {arXiv:2102.01703 [gr-qc]} \BibitemShut
  {NoStop}%
\bibitem [{\citenamefont {Rosa}\ \emph {et~al.}(2022)\citenamefont {Rosa},
  \citenamefont {Garcia}, \citenamefont {Vincent},\ and\ \citenamefont
  {Cardoso}}]{Rosa:2022toh}%
  \BibitemOpen
  \bibfield  {author} {\bibinfo {author} {\bibfnamefont {J.~a.~L.}\
  \bibnamefont {Rosa}}, \bibinfo {author} {\bibfnamefont {P.}~\bibnamefont
  {Garcia}}, \bibinfo {author} {\bibfnamefont {F.~H.}\ \bibnamefont {Vincent}},
  \ and\ \bibinfo {author} {\bibfnamefont {V.}~\bibnamefont {Cardoso}},\ }\href
  {\doibase 10.1103/PhysRevD.106.044031} {\bibfield  {journal} {\bibinfo
  {journal} {Phys. Rev. D}\ }\textbf {\bibinfo {volume} {106}},\ \bibinfo
  {pages} {044031} (\bibinfo {year} {2022})},\ \Eprint
  {http://arxiv.org/abs/2205.11541} {arXiv:2205.11541 [gr-qc]} \BibitemShut
  {NoStop}%
\bibitem [{\citenamefont {Rosa}\ and\ \citenamefont
  {Rubiera-Garcia}(2022)}]{Rosa:2022tfv}%
  \BibitemOpen
  \bibfield  {author} {\bibinfo {author} {\bibfnamefont {J.~a.~L.}\
  \bibnamefont {Rosa}}\ and\ \bibinfo {author} {\bibfnamefont {D.}~\bibnamefont
  {Rubiera-Garcia}},\ }\href {\doibase 10.1103/PhysRevD.106.084004} {\bibfield
  {journal} {\bibinfo  {journal} {Phys. Rev. D}\ }\textbf {\bibinfo {volume}
  {106}},\ \bibinfo {pages} {084004} (\bibinfo {year} {2022})},\ \Eprint
  {http://arxiv.org/abs/2204.12949} {arXiv:2204.12949 [gr-qc]} \BibitemShut
  {NoStop}%
\bibitem [{\citenamefont {Rosa}\ \emph {et~al.}(2023)\citenamefont {Rosa},
  \citenamefont {Macedo},\ and\ \citenamefont {Rubiera-Garcia}}]{Rosa:2023qcv}%
  \BibitemOpen
  \bibfield  {author} {\bibinfo {author} {\bibfnamefont {J.~a.~L.}\
  \bibnamefont {Rosa}}, \bibinfo {author} {\bibfnamefont {C.~F.~B.}\
  \bibnamefont {Macedo}}, \ and\ \bibinfo {author} {\bibfnamefont
  {D.}~\bibnamefont {Rubiera-Garcia}},\ }\href {\doibase
  10.1103/PhysRevD.108.044021} {\bibfield  {journal} {\bibinfo  {journal}
  {Phys. Rev. D}\ }\textbf {\bibinfo {volume} {108}},\ \bibinfo {pages}
  {044021} (\bibinfo {year} {2023})},\ \Eprint
  {http://arxiv.org/abs/2303.17296} {arXiv:2303.17296 [gr-qc]} \BibitemShut
  {NoStop}%
\bibitem [{\citenamefont {Jetzer}\ and\ \citenamefont {{Van Der
  Bij}}(1989)}]{JETZER1989341}%
  \BibitemOpen
  \bibfield  {author} {\bibinfo {author} {\bibfnamefont {P.}~\bibnamefont
  {Jetzer}}\ and\ \bibinfo {author} {\bibfnamefont {J.}~\bibnamefont {{Van Der
  Bij}}},\ }\href {\doibase https://doi.org/10.1016/0370-2693(89)90941-6}
  {\bibfield  {journal} {\bibinfo  {journal} {Physics Letters B}\ }\textbf
  {\bibinfo {volume} {227}},\ \bibinfo {pages} {341} (\bibinfo {year}
  {1989})}\BibitemShut {NoStop}%
\bibitem [{\citenamefont {Buchel}\ \emph {et~al.}(2013)\citenamefont {Buchel},
  \citenamefont {Liebling},\ and\ \citenamefont {Lehner}}]{Buchel:2013uba}%
  \BibitemOpen
  \bibfield  {author} {\bibinfo {author} {\bibfnamefont {A.}~\bibnamefont
  {Buchel}}, \bibinfo {author} {\bibfnamefont {S.~L.}\ \bibnamefont
  {Liebling}}, \ and\ \bibinfo {author} {\bibfnamefont {L.}~\bibnamefont
  {Lehner}},\ }\href {\doibase 10.1103/PhysRevD.87.123006} {\bibfield
  {journal} {\bibinfo  {journal} {Phys. Rev. D}\ }\textbf {\bibinfo {volume}
  {87}},\ \bibinfo {pages} {123006} (\bibinfo {year} {2013})},\ \Eprint
  {http://arxiv.org/abs/1304.4166} {arXiv:1304.4166 [gr-qc]} \BibitemShut
  {NoStop}%
\bibitem [{\citenamefont {Buchel}(2015)}]{Buchel:2015rwa}%
  \BibitemOpen
  \bibfield  {author} {\bibinfo {author} {\bibfnamefont {A.}~\bibnamefont
  {Buchel}},\ }\href@noop {} {\  (\bibinfo {year} {2015})},\ \Eprint
  {http://arxiv.org/abs/1510.08415} {arXiv:1510.08415 [hep-th]} \BibitemShut
  {NoStop}%
\bibitem [{\citenamefont {Buchel}\ and\ \citenamefont
  {Buchel}(2015)}]{Buchel:2015sma}%
  \BibitemOpen
  \bibfield  {author} {\bibinfo {author} {\bibfnamefont {A.}~\bibnamefont
  {Buchel}}\ and\ \bibinfo {author} {\bibfnamefont {M.}~\bibnamefont
  {Buchel}},\ }\href@noop {} {\  (\bibinfo {year} {2015})},\ \Eprint
  {http://arxiv.org/abs/1509.00774} {arXiv:1509.00774 [hep-th]} \BibitemShut
  {NoStop}%
\bibitem [{\citenamefont {Milekhin}\ and\ \citenamefont
  {Sukhov}(2023)}]{Milekhin:2023was}%
  \BibitemOpen
  \bibfield  {author} {\bibinfo {author} {\bibfnamefont {A.}~\bibnamefont
  {Milekhin}}\ and\ \bibinfo {author} {\bibfnamefont {N.}~\bibnamefont
  {Sukhov}},\ }\href@noop {} {\  (\bibinfo {year} {2023})},\ \Eprint
  {http://arxiv.org/abs/2307.11348} {arXiv:2307.11348 [hep-th]} \BibitemShut
  {NoStop}%
\end{thebibliography}%

\end{document}
%